\documentclass[smallextended]{svjour3}%
\usepackage{amsfonts}
\usepackage{amsmath}
\usepackage{amssymb}
\usepackage{graphicx}
\usepackage{verbatim}
\usepackage{fullpage}%
\providecommand{\U}[1]{\protect\rule{.1in}{.1in}}
\smartqed
\journalname{Quantum Information Processing}
\begin{document}

\title{Quantum Convolutional Coding with Shared Entanglement: General Structure}
\author{Mark M. Wilde
\and Todd A. Brun}
\date{Received: \today / Accepted: }
\institute{Todd Brun is a professor with the Center for Quantum Information Science and Technology and the
Communication Sciences Institute of the Ming Hsieh Department of Electrical
Engineering at the University of Southern California, Los Angeles, California
90089 USA. Mark M. Wilde was a Ph.D. student there at the completion of this research and is now
a postdoctoral fellow in the School of Computer Science, McGill University, Montreal, Quebec, Canada (E-mail: mark.wilde@mcgill.ca; tbrun@usc.edu).}
\maketitle

\begin{abstract}
We present a general theory of entanglement-assisted quantum convolutional
coding. The codes have a convolutional or memory structure, they assume that
the sender and receiver share noiseless entanglement prior to quantum
communication, and they are not restricted to possess the
Calderbank-Shor-Steane structure as in previous work. We provide two
significant advances for quantum convolutional coding theory. We first show
how to \textquotedblleft expand\textquotedblright\ a given set of quantum
convolutional generators. This expansion step acts as a preprocessor for a
polynomial symplectic Gram-Schmidt orthogonalization procedure that simplifies
the commutation relations of the expanded generators to be the same as those
of entangled Bell states (ebits)\ and ancilla qubits. The above two steps
produce a set of generators with equivalent error-correcting properties to
those of the original generators. We then demonstrate how to perform online
encoding and decoding for a stream of information qubits, halves of ebits, and
ancilla qubits. The upshot of our theory is that the quantum code designer can
engineer quantum convolutional codes with desirable error-correcting
properties without having to worry about the commutation relations of these
generators. \PACS{03.67.-a \and 03.67.Pp}

\end{abstract}

%\begin{IEEEkeywords}
%quantum convolutional codes, entanglement-assisted quantum convolutional codes, quantum
%information theory, entanglement-assisted quantum codes
%\end{IEEEkeywords}

\section{Introduction}

The theory of quantum convolutional coding has recently emerged as a powerful
means for protecting a stream of quantum information from the negative effects
of a noisy quantum communication channel
\cite{PhysRevLett.91.177902,arxiv2004olliv,isit2006grassl,ieee2006grassl,ieee2007grassl,isit2005forney,ieee2007forney,cwit2007aly,arx2007aly,arx2007wildeCED,arx2007wildeEAQCC,arx2008wildeUQCC}%
. Quantum convolutional coding theory combined with other coding techniques
may be one step along the way to finding quantum error-correcting codes that
approach the capacity of a noisy quantum communication channel for sending
quantum information
\cite{PhysRevA.55.1613,capacity2002shor,ieee2005dev,qcap2008first,qcap2008second,qcap2008third,qcap2008fourth}%
. Poulin \textit{et al}.~have recently incorporated some of this
well-developed theory of quantum convolutional coding into a theory of quantum
serial-turbo coding \cite{arx2007poulin}\ with the goal of designing quantum
codes that come close to achieving capacity.

Quantum convolutional codes have numerous benefits. The periodic structure of
their encoding and decoding circuits ensures a low complexity for encoding and
decoding while also providing higher performance than a block code with
equivalent encoding complexity \cite{ieee2007forney}. The encoding and
decoding circuits have the property that the sender Alice and the receiver Bob
can respectively send and receive qubits in an \textquotedblleft
online\textquotedblright\ fashion. Alice can encode an arbitrary number of
information qubits without worrying beforehand how many she may want to send
over the quantum communication channel.

A seminal paper on quantum error correction showed how to produce quantum
error-correcting codes from classical linear binary or quaternary block codes
\cite{ieee1998calderbank}. Importing binary codes to construct quantum codes
is now known as the Calderbank-Shor-Steane (CSS)\ construction
\cite{book2000mikeandike}. Importing classical codes is useful because quantum
code designers can utilize classical block codes with high performance to
construct quantum codes with high performance. The only problem with this
technique is that the imported classical block codes have to satisfy a
restrictive dual-containing condition so that the operators in the resulting
quantum code commute with one another.

Brun, Devetak, and Hsieh alleviated the dual-containing constraint by assuming
that the sender and receiver share noiseless entanglement in the
entanglement-assisted stabilizer formalism \cite{science2006brun,arx2006brun}.
The entanglement-assisted stabilizer formalism allows quantum code designers
to produce a quantum code from an arbitrary classical binary or quaternary
block code by incorporating shared entanglement. In a different context,
Forney \textit{et al}.~showed how to produce quantum convolutional codes from
classical convolutional codes by extending the ideas from the
CSS\ construction to the convolutional setting \cite{ieee2007forney}; but
again, these imported classical convolutional codes have to satisfy the
restrictive dual-containing constraint in order to form valid quantum codes
(the dual-containing constraint is actually quite a bit more restrictive in
the convolutional case). The current authors followed these two developments
by detailing entanglement-assisted quantum convolutional codes
\cite{arx2007wildeEAQCC}\ that have the CSS\ structure. We demonstrated in
Ref. \cite{arx2007wildeEAQCC}\ how to produce an entanglement-assisted quantum
convolutional code from two \textit{arbitrary} classical binary convolutional
codes---these imported codes do not need to satisfy a dual-containing
constraint. The benefit of all these entanglement-assisted techniques is that
we can use high-performance classical codes for correction of quantum errors.

In this paper, we show how to encode and decode an entanglement-assisted
quantum convolutional code that does not necessarily have the CSS\ structure.
The methods in this work represent a significant extension of our previous
work on CSS\ entanglement-assisted quantum convolutional codes
\cite{arx2007wildeEAQCC}. In particular, we develop a set of techniques that
make the commutation relations for the generators of a general code the same
as those of entangled qubits (ebits) and ancilla qubits. This procedure first
\textquotedblleft expands\textquotedblright\ the check matrix for a quantum
convolutional code and applies an extended version of the symplectic
Gram-Schmidt orthogonalization procedure \cite{arx2006brun}\ that incorporates
the binary polynomials representing quantum convolutional codes. We then show
how to encode a stream of information qubits, ancilla qubits, and ebits to
have the error-correcting properties of the desired code. We follow by
detailing the decoding circuit that Bob employs at the receiving end of the
channel. The algorithms for encoding and decoding use similar techniques to
those outlined in Refs.~\cite{isit2006grassl,grassl2006itw,arx2007wildeEAQCC}.
The encoding circuits incorporate both finite-depth and infinite-depth
\cite{arx2007wildeEAQCC}\ operations and the decoding circuits incorporate
finite-depth operations only. One benefit of the techniques developed in this
paper is that the quantum code designer can produce an entanglement-assisted
quantum convolutional code from a classical quaternary convolutional code.
More generally, quantum convolutional code designers now have the freedom to
design quantum convolutional codes with desirable error-correcting properties
without having to search for codes that satisfy a commutativity constraint.

An important practical issue in the design of entanglement-assisted quantum
convolutional codes is the characterization of performance in terms of some
parameter like code distance or distance spectrum~\cite{arx2007poulin}. The
focus of this paper is on the construction and design of encoding and decoding
circuits for non-CSS entanglement-assisted quantum convolutional codes---not
on the performance of these codes. We take this approach because
entanglement-assisted codes produced from classical codes inherit distance and
error-correcting properties from the parent classical codes. If the
entanglement-assisted code does not come from a classical code, then one can
analyze distance of the code in the standard way as the minimum weight element
of the normalizer modulo the stabilizer \cite{book2000mikeandike} or with the
distance spectrum parameter of Poulin \textit{et al}.~\cite{arx2007poulin},
but an analysis of this sort is beyond the scope of the present paper.

We structure our work as follows. Section~\ref{sec:review-qcc} reviews the
general definition of a quantum convolutional code and the matrix of binary
polynomials that represents it. It also discusses the shifted symplectic
product that captures the commutation relations of Pauli sequences.
Section~\ref{sec:review-EAQCC} reviews the theory of entanglement-assisted
quantum block coding. In Section \ref{sec:expand}, we present an example that
demonstrates how to expand the check matrix of a set of quantum convolutional
generators and then show how to expand an arbitrary check matrix. We show in
Section~\ref{sec:GS}\ how to compute the symplectic Gram-Schmidt
orthogonalization procedure for the example in Section~\ref{sec:expand} and
then generalize this procedure to work for an arbitrary set of quantum
convolutional generators. The Gram-Schmidt orthogonalization technique reduces
the quantum check matrix to have the same commutation relations as a set of
ebits and ancilla qubits. This technique is essential for determining an
encoding circuit that encodes a stream of information qubits, ancilla qubits,
and ebits. Section \ref{sec:encode-decode}\ gives the algorithms for computing
the encoding and decoding circuits for an arbitrary entanglement-assisted
quantum convolutional code. We present a detailed example in
Section~\ref{sec:examples}\ that illustrates all of the procedures in this
article. We finish with a discussion of the practical issues involved in using
these entanglement-assisted quantum convolutional codes and present some
concluding remarks.

\section{Review of Quantum Convolutional Codes}

\label{sec:review-qcc}We first review the theory of quantum convolutional
coding. The reader can find more detailed reviews in
Refs.~\cite{ieee2007forney,arx2007wildeCED,arx2007wildeEAQCC}. A quantum
convolutional code admits a representation in terms of Pauli
sequences---infinite tensor products of Pauli matrices. We demonstrate how to
map these Pauli sequences to a finite-dimensional binary polynomial matrix and
call this mapping the Pauli-to-binary (P2B)\ isomorphism. It is more
straightforward to work with binary polynomial matrices rather than Pauli
sequences because we can employ linear-algebraic techniques.

We state some basic definitions and notation before proceeding with the
definition of a quantum convolutional code. A Pauli sequence $\mathbf{A}$ is a
countably infinite tensor product of Pauli matrices:%
\[
\mathbf{A}=%
%TCIMACRO{\dbigotimes \limits_{i=0}^{\infty}}%
%BeginExpansion
{\displaystyle\bigotimes\limits_{i=0}^{\infty}}
%EndExpansion
A_{i},
\]
where each $A_{i}$ is a matrix in the Pauli group $\Pi=\left\{
I,X,Y,Z\right\}  $. Let $\Pi^{\mathbb{Z}^{+}}$ denote the set of all Pauli
sequences. A finite-weight Pauli sequence is one in which the operators $X$,
$Y$, and $Z$ act on a finite number of qubits and the identity operator acts
on the countably infinite remaining qubits. An infinite-weight sequence is one
that is not finite-weight.

\begin{definition}
A rate-$k/n$ quantum convolutional code admits a representation by a basic set
$\mathcal{G}_{0}$ of $n-k$ generators and all of their $n$-qubit shifts. For a
quantum convolutional code, the generators commute with each other and all of
the $n$-qubit shifts of themselves and the other generators. This
commutativity requirement is necessary for the same reason that standard
stabilizer codes require it \cite{book2000mikeandike}. The generators in the
basic set $\mathcal{G}_{0}$ are as follows:%
\[
\mathcal{G}_{0}=\left\{  \mathbf{G}_{i}\in\Pi^{\mathbb{Z}^{+}}:1\leq i\leq
n-k\right\}  .
\]
A frame of the code consists of $n$ qubits so that the frame size is $n$. The
definition of a quantum convolutional code as $n$-qubit shifts of the basic
set $\mathcal{G}_{0}$\ is what gives the code its periodic structure.
\end{definition}

We do not discuss the detailed operation of a quantum convolutional code in
this article. Please consult
Refs.~\cite{ieee2007forney,arx2007wildeCED,arx2007wildeEAQCC} for a discussion
of a quantum convolutional code's operation and see Figure~3 of
Ref.~\cite{arx2007wildeCED} for a depiction of its operation. The current
article addresses the issues of quantum convolutional code design.

We define the phase-free Pauli group $\left[  \Pi^{\mathbb{Z}}\right]  $\ on a
sequence of qubits. The delay transform $D$ shifts a Pauli sequence to the
right by $n$ qubits \cite{ieee2007forney}. Let $\Pi^{\mathbb{Z}}$ denote the
set of all countably infinite Pauli sequences. The set $\Pi^{\mathbb{Z}}$ is
equivalent to the set of all $n$-qubit shifts of arbitrary Pauli operators:%
\begin{equation}
\Pi^{\mathbb{Z}}=\left\{
%TCIMACRO{\tprod \limits_{i\in\mathbb{Z}}}%
%BeginExpansion
{\textstyle\prod\limits_{i\in\mathbb{Z}}}
%EndExpansion
D^{i}\left(  A_{i}\right)  :A_{i}\in\Pi^{n}\right\}  ,
\end{equation}
where $\Pi^{n}$ is the Pauli group over $n$ qubits. We remark that
$D^{i}\left(  A_{i}\right)  =D^{i}\left(  A_{i}\otimes I^{\otimes\infty
}\right)  $. We make this same abuse of notation in what follows. We can
define the equivalence class $\left[  \Pi^{\mathbb{Z}}\right]  $\ of
phase-free Pauli sequences:%
\begin{equation}
\left[  \Pi^{\mathbb{Z}}\right]  =\left\{  \beta\mathbf{A\ }|\ \mathbf{A}%
\in\Pi^{\mathbb{Z}},\beta\in\mathbb{C},\left\vert \beta\right\vert =1\right\}
.
\end{equation}

We develop a relation between binary polynomials and Pauli sequences that is
useful for representing the shifting nature of quantum convolutional codes. We
call this map the Pauli-to-binary (P2B) isomorphism. A quantum convolutional
code in general consists of generators with $n$ qubits per frame. One can
obtain the full set of generators of the quantum convolutional code by
shifting the basic generator set by integer multiples of the frame size $n$
\cite{ieee2007forney}. Let the delay transform $D$ shift a Pauli sequence to
the right by an arbitrary integer $n$. Consider a $2n$-dimensional vector
$\mathbf{u}\left(  D\right)  $ of binary polynomials where $\mathbf{u}\left(
D\right)  \in\left(  \mathbb{Z}_{2}\left(  D\right)  \right)  ^{2n}$. Let us
write $\mathbf{u}\left(  D\right)  $\ as follows:%
\begin{align*}
\mathbf{u}\left(  D\right)   &  =\left(  \mathbf{z}\left(  D\right)
|\mathbf{x}\left(  D\right)  \right)  ,\\
&  =\left(
\begin{array}
[c]{ccc}%
z_{1}\left(  D\right)  & \cdots & z_{n}\left(  D\right)
\end{array}
|%
\begin{array}
[c]{ccc}%
x_{1}\left(  D\right)  & \cdots & x_{n}\left(  D\right)
\end{array}
\right)  ,
\end{align*}
where $\mathbf{z}\left(  D\right)  ,\mathbf{x}\left(  D\right)  \in\left(
\mathbb{Z}_{2}\left(  D\right)  \right)  ^{n}$. Suppose%
\[
z_{i}\left(  D\right)  =\sum_{j}z_{i,j}D^{j},\ \ \ \ x_{i}\left(  D\right)
=\sum_{j}x_{i,j}D^{j}.
\]
Define a map $\mathbf{N}:\left(  \mathbb{Z}_{2}\left(  D\right)  \right)
^{2n}\rightarrow\Pi^{\mathbb{Z}}$:%
\begin{equation}
\mathbf{N}\left(  \mathbf{u}\left(  D\right)  \right)  =%
%TCIMACRO{\tprod \limits_{j}}%
%BeginExpansion
{\textstyle\prod\limits_{j}}
%EndExpansion
D^{j}\left(  Z^{z_{1,j}}X^{x_{1,j}}\right)  D^{j}\left(  I\otimes Z^{z_{2,j}%
}X^{x_{2,j}}\right)  \cdots D^{j}\left(  I^{\otimes n-1}\otimes Z^{z_{n,j}%
}X^{x_{n,j}}\right)  .\nonumber
\end{equation}
$\mathbf{N}$ is equivalent to the following map (up to a global phase)%
\begin{equation}
\mathbf{N}\left(  \mathbf{u}\left(  D\right)  \right)  =N\left(  u_{1}\left(
D\right)  \right)  \left(  I\otimes N\left(  u_{2}\left(  D\right)  \right)
\right)  \cdots\left(  I^{\otimes n-1}\otimes N\left(  u_{n}\left(  D\right)
\right)  \right)  ,\nonumber
\end{equation}
where $u_{i}\left(  D\right)  =\left(  z_{i}\left(  D\right)  |x_{i}\left(
D\right)  \right)  $. Suppose $\mathbf{v}\left(  D\right)  =\left(
\mathbf{z}^{\prime}\left(  D\right)  |\mathbf{x}^{\prime}\left(  D\right)
\right)  $, where $\mathbf{v}\left(  D\right)  \in\left(  \mathbb{Z}%
_{2}\left(  D\right)  \right)  ^{2n}$. The map $\mathbf{N}$\ induces an
isomorphism
\[
\left[  \mathbf{N}\right]  :\left(  \mathbb{Z}_{2}\left(  D\right)  \right)
^{2n}\rightarrow\left[  \Pi^{\mathbb{Z}}\right]
\]
because addition of binary polynomials is equivalent to multiplication of
Pauli elements up to a global phase:
\[
\left[  \mathbf{N}\left(  \mathbf{u}\left(  D\right)  +\mathbf{v}\left(
D\right)  \right)  \right]  =\left[  \mathbf{N}\left(  \mathbf{u}\left(
D\right)  \right)  \right]  \left[  \mathbf{N}\left(  \mathbf{v}\left(
D\right)  \right)  \right]  .
\]

We illustrate the P2B isomorphism by means of an example. Consider the
following generator for a quantum convolutional code:%
\[
\cdots\left\vert
\begin{array}
[c]{c}%
III
\end{array}
\right.  \left\vert
\begin{array}
[c]{c}%
XXX
\end{array}
\right\vert
\begin{array}
[c]{c}%
XZY
\end{array}
\left\vert
\begin{array}
[c]{c}%
III
\end{array}
\right\vert \cdots
\]
We obtained the above generator by modifying generators for a code from
Ref.~\cite{ieee2007forney}. The vertical bars indicate that the frame size is
three qubits and every three-qubit shift of the above generator yields another
stabilizer generator for the code. The above generator forms a valid quantum
convolutional code because it commutes with all of its three-qubit shifts. For
now, we are not concerned with the error-correcting properties of the above
generator---we are merely interested in illustrating the principle of the P2B isomorphism.

The above generator has the following representation as a binary polynomial
matrix:%
\[
\left[  \left.
\begin{array}
[c]{ccc}%
0 & D & D
\end{array}
\right\vert
\begin{array}
[c]{ccc}%
1+D & 1 & 1+D
\end{array}
\right]  .
\]
The matrix on the right is the \textquotedblleft X\textquotedblright\ matrix
and the matrix on the left is the \textquotedblleft Z\textquotedblright%
\ matrix. The vertical bar in the above matrix serves to distinguish between
the \textquotedblleft X\textquotedblright\ and \textquotedblleft
Z\textquotedblright\ matrices. Each \textquotedblleft X\textquotedblright%
\ operator in the first frame of the code gives a \textquotedblleft%
1\textquotedblright\ entry in the \textquotedblleft X\textquotedblright%
\ matrix. There are no \textquotedblleft1\textquotedblright\ entries in the
\textquotedblleft Z\textquotedblright\ matrix because there are no
\textquotedblleft Z\textquotedblright\ operators in the first frame of the
code. Each \textquotedblleft X\textquotedblright\ and \textquotedblleft
Z\textquotedblright\ operator in the second frame of the code gives a
\textquotedblleft$D$\textquotedblright\ entry in the respective
\textquotedblleft X\textquotedblright\ and \textquotedblleft
Z\textquotedblright\ matrices. The \textquotedblleft Y\textquotedblright%
\ entry in the second frame contributes a \textquotedblleft$D$%
\textquotedblright\ entry to both the \textquotedblleft X\textquotedblright%
\ and \textquotedblleft Z\textquotedblright\ matrix.

In general, we represent a rate-$k/n$ quantum convolutional code with an
$\left(  n-k\right)  \times2n$-dimensional quantum check matrix $H\left(
D\right)  $ whose entries are binary polynomials where%
\[
H\left(  D\right)  =\left[  \left.
\begin{array}
[c]{c}%
Z\left(  D\right)
\end{array}
\right\vert
\begin{array}
[c]{c}%
X\left(  D\right)
\end{array}
\right]  .
\]

The shifted symplectic product $\odot$\ captures the commutations relations of
any two generators \cite{arx2007wildeCED,arx2007wildeEAQCC}. Suppose that
$h_{1}\left(  D\right)  =\left[  \left.
\begin{array}
[c]{c}%
z_{1}\left(  D\right)
\end{array}
\right\vert
\begin{array}
[c]{c}%
x_{1}\left(  D\right)
\end{array}
\right]  $ and $h_{2}\left(  D\right)  =\left[  \left.
\begin{array}
[c]{c}%
z_{2}\left(  D\right)
\end{array}
\right\vert
\begin{array}
[c]{c}%
x_{2}\left(  D\right)
\end{array}
\right]  $ are the first and second respective rows of check matrix $H\left(
D\right)  $. The shifted symplectic product $\left(  h_{1}\odot h_{2}\right)
\left(  D\right)  $ is a polynomial equal to the following expression:%
\[
\left(  h_{1}\odot h_{2}\right)  \left(  D\right)  =z_{1}\left(  D\right)
\cdot x_{2}\left(  D^{-1}\right)  +x_{1}\left(  D\right)  \cdot z_{2}\left(
D^{-1}\right)  ,
\]
where \textquotedblleft$\cdot$\textquotedblright\ represents the standard
inner product between two vectors and addition is binary. Please note that we
have slightly changed the convention for the shifted symplectic product from
that used in Ref's.~\cite{arx2007wildeCED,arx2007wildeEAQCC} in order to agree
with the convention in Ref's.~\cite{isit2006grassl,ieee2007grassl}. The
shifted symplectic products $\left(  h_{1}\odot h_{1}\right)  \left(
D\right)  $, $\left(  h_{2}\odot h_{2}\right)  \left(  D\right)  $, and
$\left(  h_{1}\odot h_{2}\right)  \left(  D\right)  $ are equal to zero for in
the matrix $H(D)$ because these generators are part of a valid quantum
convolutional code and therefore commute with each other and with all shifts
of themselves and each other. For a general set of generators, this condition
does not necessarily have to hold. We spend much of our effort in this article
developing techniques to modify the commutation relations of a general set of generators.

The following matrix $\Omega\left(  D\right)  $ captures the commutation
relations of the generators in $H\left(  D\right)  $:%
\[
\Omega\left(  D\right)  =Z\left(  D\right)  X^{T}\left(  D^{-1}\right)
+X\left(  D\right)  Z^{T}\left(  D^{-1}\right)  .
\]
The reader can verify that the matrix elements $\left[  \Omega\left(
D\right)  \right]  _{ij}$ of $\Omega\left(  D\right)  $ are as follows:%
\[
\left[  \Omega\left(  D\right)  \right]  _{ij}=\left(  h_{i}\odot
h_{j}\right)  \left(  D\right)  ,
\]
where $\odot$ denotes the shifted symplectic product. The matrix
$\Omega\left(  D\right)  $ obeys the symmetry:\ $\Omega\left(  D\right)
=\Omega^{T}\left(  D^{-1}\right)  $. We call the matrix $\Omega\left(
D\right)  $ the \textit{shifted symplectic product matrix} because it encodes
all of the shifted symplectic products or the commutation relations of the
code. This matrix is equal to the null matrix for a valid quantum
convolutional code because all generators commute with themselves and with all
$n$-qubit shifts of themselves and each other. For a general set of
generators, $\Omega\left(  D\right)  $ is not necessarily equal to the null matrix.

We comment briefly on row operations. We can add one row of the check matrix
$H\left(  D\right)  $ to another row in it without changing the
error-correcting properties of the code. This property holds because the code
is additive. We can even premultiply check matrix $H\left(  D\right)  $\ by an
arbitrary matrix $R\left(  D\right)  $ with rational polynomial entries
without changing the error-correcting properties of the code. This result
follows from the classical theory of convolutional coding \cite{book1999conv}.
Let $H^{\prime}\left(  D\right)  $ denote the resulting check matrix where
$H^{\prime}\left(  D\right)  =R\left(  D\right)  H\left(  D\right)  $. The
resulting effect on the shifted symplectic product matrix $\Omega\left(
D\right)  $ is to change it to another shifted symplectic product matrix
$\Omega^{\prime}\left(  D\right)  $ related to $\Omega\left(  D\right)  $ by%
\begin{equation}
\Omega^{\prime}\left(  D\right)  =R\left(  D\right)  \Omega\left(  D\right)
R^{T}\left(  D^{-1}\right)  . \label{eq:symp-row-op}%
\end{equation}
Row operations do not change the commutation relations of a valid quantum
convolutional code because its shifted symplectic product matrix is equal to
the null matrix. But row operations do change the commutation relations of a
set of generators whose corresponding shifted symplectic product matrix is not
equal to the null matrix. This ability to change the commutation relations
through row operations is crucial for constructing entanglement-assisted
quantum convolutional codes from an arbitrary set of generators. We use
entanglement to resolve any anticommutativity in the generators.

\section{Review of Entanglement-Assisted Quantum Block Codes}

\label{sec:review-EAQCC}We briefly review the theory of entanglement-assisted
quantum coding for block quantum codes. The entanglement-assisted stabilizer
formalism subsumed the stabilizer formalism
\cite{thesis97gottesman,ieee1998calderbank} by assuming that the sender and
receiver in a quantum communication protocol can share entanglement
\cite{arx2006brun,science2006brun}. This technique of incorporating shared
entanglement allows the sender and receiver to import arbitrary classical
codes for use in quantum error correction.

Consider the ebit or Bell state $\left\vert \Phi^{+}\right\rangle ^{AB}$
shared between the sender Alice $A$ and the receiver Bob $B$:%
\[
\left\vert \Phi^{+}\right\rangle ^{AB}=\frac{\left\vert 0\right\rangle
^{A}\left\vert 0\right\rangle ^{B}+\left\vert 1\right\rangle ^{A}\left\vert
1\right\rangle ^{B}}{\sqrt{2}}.
\]
Two operators that stabilize this state are $X^{A}X^{B}$ and $Z^{A}Z^{B}$.
These two operators commute,%
\[
\left[  X^{A}X^{B},Z^{A}Z^{B}\right]  =0,
\]
but consider that the local operators operating only on either party
anticommute,%
\[
\left\{  X^{A},Z^{A}\right\}  =\left\{  X^{B},Z^{B}\right\}  =0.
\]

The above commutation relations hint at a way that we can resolve
anticommutativity in a set of generators. Suppose that we have two generators
in a code that anticommute. We can resolve the anticommutativity in the two
generators by adding an extra $X$ or $Z$ operator to the generators on the
receiver's side, at the cost of an ebit of entanglement. We can find an
encoding circuit starting from our unencoded information qubits and one ebit
to encode the generators for the code. Detailed algorithms exist to find an
encoding circuit---please see
Refs.~\cite{arx2006brun,science2006brun,arx2007wildeCED,prep2007shaw}. These
algorithms can also handle a set of generators that have more complicated
commutation relations. It turns out that these algorithms use the optimal
number of ebits~for a given set of generators \cite{arx2008wildeOEA}.

Let us write the local operators $X^{A}$ and $Z^{A}$ as respective symplectic
vectors $g_{A}$ and $h_{A}$:%
\[
g_{A}=\left[  \left.
\begin{array}
[c]{c}%
0
\end{array}
\right\vert
\begin{array}
[c]{c}%
1
\end{array}
\right]  ,\ \ \ \ \ \ h_{A}=\left[  \left.
\begin{array}
[c]{c}%
1
\end{array}
\right\vert
\begin{array}
[c]{c}%
0
\end{array}
\right]  .
\]
The symplectic product matrix $\Omega$\ of these two symplectic vectors is as
follows:%
\begin{equation}
\Omega=\left[
\begin{array}
[c]{cc}%
0 & 1\\
1 & 0
\end{array}
\right]  . \label{eq:J-matrix}%
\end{equation}
The above symplectic product matrix is so special for the purposes of this
article that we give it the name $J\equiv\Omega$. It means that two generators
have commutation relations that are equivalent to half of an ebit that the
sender possesses.

A general set of generators for a quantum block code can have complicated
commutation relations. There exists a symplectic Gram-Schmidt
orthogonalization procedure that simplifies the commutation relations
\cite{arx2006brun,science2006brun,arx2007wildeCED,prep2007shaw}. Specifically,
the algorithm performs row operations that do not affect the code's
error-correcting properties and thus gives a set of generators that form an
equivalent code. For a given set of $n-k$ generators, the commutation
relations for the generators resulting from the algorithm have a special form.
Their symplectic product matrix $\Omega$\ has the standard form:%
\begin{equation}
\Omega=%
%TCIMACRO{\dbigoplus \limits_{i=1}^{c}}%
%BeginExpansion
{\displaystyle\bigoplus\limits_{i=1}^{c}}
%EndExpansion
J\oplus%
%TCIMACRO{\dbigoplus \limits_{j=1}^{a}}%
%BeginExpansion
{\displaystyle\bigoplus\limits_{j=1}^{a}}
%EndExpansion
\left[  0\right]  , \label{eq:standard-symp-form}%
\end{equation}
where the large and small $\oplus$ correspond to the direct sum operation and
$\left[  0\right]  $ is the one-element null matrix. The standard form above
implies that the commutation relations of the $n-k$ generators are equivalent
to those of $c$ halves of ebits and $a$ ancilla qubits where $a=n-k-2c$. It is
straightforward to find an encoding circuit that encodes $k+c$ information
qubits, $a$ ancilla qubits, and $c$ halves of ebits once we have reduced the
commutation relations of the generators to have the above standard form.

The focus of the next two sections is to elucidate techniques for reducing a
set of quantum convolutional generators to have the standard commutation
relations given above. The difference between the current techniques and the
former techniques
\cite{arx2006brun,science2006brun,arx2007wildeCED,prep2007shaw} is that the
current techniques operate on generators that have a convolutional form rather
than a block form. The next section shows how to expand a set of quantum
convolutional generators to simplify their commutation relations. The section
following the next outlines a symplectic Gram-Schmidt orthogonalization
algorithm that uses binary polynomial operations to reduce the commutation
relations of the expanded generators to have the standard form.

\section{The Expansion of Quantum Convolutional Generators}

\label{sec:expand}We begin this section by demonstrating how to expand a
particular generator that we eventually incorporate in an
entanglement-assisted quantum convolutional code. We later generalize this
example and the expansion technique to an arbitrary set of generators. This
technique is important for determining how to utilize entanglement in the form
of ebits in a quantum convolutional code.

\subsection{Example of the Expansion}

Let us first suppose that we have one convolutional generator:%
\begin{equation}
\cdots\left\vert
\begin{array}
[c]{c}%
I
\end{array}
\right.  \left\vert
\begin{array}
[c]{c}%
X
\end{array}
\right\vert
\begin{array}
[c]{c}%
Z
\end{array}
\left\vert
\begin{array}
[c]{c}%
I
\end{array}
\right\vert \cdots\label{eq:pauli-conv-simple}%
\end{equation}
This generator does not yet represent a valid quantum convolutional code
because it anticommutes with a shift of itself by one qubit to the left or to
the right. We are not concerned with the error-correcting properties of this
generator but merely want to illustrate the technique of expanding it.

Let us for now consider a block version of this code that operates on six
physical qubits with six total generators. The generators are as follows:%
\begin{equation}
\left.
\begin{array}
[c]{c}%
X\\
I\\
I\\
I\\
I\\
I
\end{array}
\right\vert
\begin{array}
[c]{c}%
Z\\
X\\
I\\
I\\
I\\
I
\end{array}
\left\vert
\begin{array}
[c]{c}%
I\\
Z\\
X\\
I\\
I\\
I
\end{array}
\right\vert
\begin{array}
[c]{c}%
I\\
I\\
Z\\
X\\
I\\
I
\end{array}
\left\vert
\begin{array}
[c]{c}%
I\\
I\\
I\\
Z\\
X\\
I
\end{array}
\right\vert
\begin{array}
[c]{c}%
I\\
I\\
I\\
I\\
Z\\
X
\end{array}
. \label{eq:pauli-block-simple}%
\end{equation}
We still use the vertical bars in the above block code to denote that the
frame size of the code is one. Observe that we can view the frame size of the
code as two without changing any of the error-correcting properties of the
block code:%
\begin{equation}
\left.
\begin{array}
[c]{cc}%
X & Z\\
I & X\\
I & I\\
I & I\\
I & I\\
I & I
\end{array}
\right\vert
\begin{array}
[c]{cc}%
I & I\\
Z & I\\
X & Z\\
I & X\\
I & I\\
I & I
\end{array}
\left\vert
\begin{array}
[c]{cc}%
I & I\\
I & I\\
I & I\\
Z & I\\
X & Z\\
I & X
\end{array}
\right.  . \label{eq:two-expanded-code}%
\end{equation}
The frame is merely a way to organize our qubits so that we send one frame at
a time over the channel after the online encoding circuit has finished
processing them. We can extend the above block code with frame size two to
have a convolutional structure with the following two convolutional
generators:%
\[
\cdots\left\vert
\begin{array}
[c]{cc}%
I & I\\
I & I
\end{array}
\right.  \left\vert
\begin{array}
[c]{cc}%
X & Z\\
I & X
\end{array}
\right\vert
\begin{array}
[c]{cc}%
I & I\\
Z & I
\end{array}
\left\vert
\begin{array}
[c]{cc}%
I & I\\
I & I
\end{array}
\right\vert \cdots
\]
The above two generators with frame size two have equivalent error-correcting
properties to the original generator in (\ref{eq:pauli-conv-simple}) by the
arguments above. We say that we have expanded the original generator by a
factor of two or that the above code is a two-expanded version of the original
generator. We can also extend the original generator in
(\ref{eq:pauli-conv-simple})\ to have a frame size of three and we require
three convolutional generators so that they have equivalent error-correcting
properties to the original generator:%
\[
\cdots\left\vert
\begin{array}
[c]{ccc}%
I & I & I\\
I & I & I\\
I & I & I
\end{array}
\right.  \left\vert
\begin{array}
[c]{ccc}%
X & Z & I\\
I & X & Z\\
I & I & X
\end{array}
\right\vert
\begin{array}
[c]{ccc}%
I & I & I\\
I & I & I\\
Z & I & I
\end{array}
\left\vert
\begin{array}
[c]{ccc}%
I & I & I\\
I & I & I\\
I & I & I
\end{array}
\right\vert \cdots
\]

The representation of the original generator in (\ref{eq:pauli-conv-simple})
as a quantum check matrix in the polynomial formalism is as follows:%
\begin{equation}
g\left(  D\right)  =\left[  \left.
\begin{array}
[c]{c}%
D
\end{array}
\right\vert
\begin{array}
[c]{c}%
1
\end{array}
\right]  . \label{eq:poly-conv-simple}%
\end{equation}
The two-expanded check matrix has the following polynomial representation:%
\begin{equation}
G_{2}\left(  D\right)  =\left[  \left.
\begin{array}
[c]{cc}%
0 & 1\\
D & 0
\end{array}
\right\vert
\begin{array}
[c]{cc}%
1 & 0\\
0 & 1
\end{array}
\right]  , \label{eq:two-exp-check}%
\end{equation}
and the three-expanded check matrix has the following polynomial
representation:%
\[
G_{3}\left(  D\right)  =\left[  \left.
\begin{array}
[c]{ccc}%
0 & 1 & 0\\
0 & 0 & 1\\
D & 0 & 0
\end{array}
\right\vert
\begin{array}
[c]{ccc}%
1 & 0 & 0\\
0 & 1 & 0\\
0 & 0 & 1
\end{array}
\right]  .
\]

An alternative method for obtaining the polynomial representation of the
two-expanded check matrix consists of two steps. We first multiply $g\left(
D\right)  $ as follows:%
\[
G_{2}^{\prime}\left(  D\right)  =\left[
\begin{array}
[c]{c}%
1\\
D
\end{array}
\right]  \left[  g\left(  D\right)  \right]  \left[
\begin{array}
[c]{cccc}%
1 & D^{-1} & 0 & 0\\
0 & 0 & 1 & D^{-1}%
\end{array}
\right]  .
\]
We then \textquotedblleft plug in\textquotedblright\ the fractional delay
operator $D^{1/2}$ and apply the flooring operation $\left\lfloor
\cdot\right\rfloor $ that nulls the coefficients of any fractional power of
\thinspace$D$:%
\[
G_{2}\left(  D\right)  =\left\lfloor G_{2}^{\prime}\left(  D^{1/2}\right)
\right\rfloor .
\]
A similar technique applies to find the check matrix of the three-expanded
matrix. We first multiply $g\left(  D\right)  $ as follows:%
\[
G_{3}\left(  D\right)  =\left[
\begin{array}
[c]{c}%
1\\
D\\
D^{2}%
\end{array}
\right]  \left[  g\left(  D\right)  \right]  \left[
\begin{array}
[c]{cccccc}%
1 & \frac{1}{D} & \frac{1}{D^{2}} & 0 & 0 & 0\\
0 & 0 & 0 & 1 & \frac{1}{D} & \frac{1}{D^{2}}%
\end{array}
\right]  .
\]
We then \textquotedblleft plug in\textquotedblright\ the fractional delay
operator $D^{1/3}$ and apply the flooring operation $\left\lfloor
\cdot\right\rfloor $:%
\[
G_{3}\left(  D\right)  =\left\lfloor G_{3}^{\prime}\left(  D^{1/3}\right)
\right\rfloor .
\]
We discuss the general method for expanding an arbitrary check matrix in the
next subsection.

\subsection{General Technique for Expansion}

We generalize the above example to determine how to expand an arbitrary
quantum check matrix by a factor of $l$. Suppose that we have an $n-k\times2n$
quantum check matrix $H\left(  D\right)  $ where%
\[
H\left(  D\right)  =\left[  \left.
\begin{array}
[c]{c}%
Z\left(  D\right)
\end{array}
\right\vert
\begin{array}
[c]{c}%
X\left(  D\right)
\end{array}
\right]  .
\]
Let $\mathbf{D}$ denote a diagonal matrix whose diagonal entries are the delay
operator $D$. We take the convention that $\mathbf{D}^{0}$ is the identity
matrix and $\mathbf{D}^{m}$ is the matrix $\mathbf{D}$ multiplied $m$ times so
that its diagonal entries are $D^{m}$. Let $R_{l}\left(  D\right)  $ and
$C_{l}\left(  D\right)  $ denote the following matrices:%
\[
R_{l}\left(  D\right)  \equiv\left[
\begin{array}
[c]{cccc}%
\mathbf{D}^{0} & \mathbf{D}^{1} & \cdots & \mathbf{D}^{l-1}%
\end{array}
\right]  ^{T},\ \ \ \ \ \ \ \ \ C_{l}\left(  D\right)  \equiv\left[
\begin{array}
[c]{cccccc}%
\mathbf{D}^{0} & \cdots & \mathbf{D}^{-\left(  l-1\right)  } & 0 & \cdots &
0\\
0 & \cdots & 0 & \mathbf{D}^{0} & \cdots & \mathbf{D}^{-\left(  l-1\right)  }%
\end{array}
\right]  ,
\]
where the diagonal $\mathbf{D}$ matrices in $R_{l}\left(  D\right)  $ and
$C_{l}\left(  D\right)  $ have respective dimensions $n-k\times n-k$ and
$n\times n$. We premultiply and postmultiply the matrix $H\left(  D\right)  $
by respective matrices $R_{l}\left(  D\right)  $ and $C_{l}\left(  D\right)
$, evaluate the resulting matrix at a fractional power $1/l$ of the delay
operator $D$, and perform the flooring operation $\left\lfloor \cdot
\right\rfloor $ to null the coefficients of any fractional power of
\thinspace$D$. The $l$-expanded check matrix $H_{l}\left(  D\right)  $ is as
follows:%
\[
H_{l}\left(  D\right)  =\left\lfloor R_{l}\left(  D^{1/l}\right)  H\left(
D^{1/l}\right)  C_{l}\left(  D^{1/l}\right)  \right\rfloor .
\]
The $l$-expanded quantum check matrix $H_{l}\left(  D\right)  $ has equivalent
error-correcting properties to the original check matrix. One can verify that
the above property holds by carrying out the above operations:%
\begin{align*}
H_{l}\left(  D\right)   &  =\left\lfloor R_{l}\left(  D^{1/l}\right)  H\left(
D^{1/l}\right)  C_{l}\left(  D^{1/l}\right)  \right\rfloor \\
&  =\left\lfloor \left[
\begin{array}
[c]{c}%
\mathbf{D}^{0}\\
\mathbf{D}^{1/l}\\
\vdots\\
\mathbf{D}^{\left(  l-1\right)  /l}%
\end{array}
\right]  \left[  \left.
\begin{array}
[c]{c}%
Z\left(  D^{1/l}\right)
\end{array}
\right\vert
\begin{array}
[c]{c}%
X\left(  D^{1/l}\right)
\end{array}
\right]  \left[
\begin{array}
[c]{cccccc}%
\mathbf{D}^{0} & \cdots & \mathbf{D}^{-\left(  l-1\right)  /l} & 0 & \cdots &
0\\
0 & \cdots & 0 & \mathbf{D}^{0} & \cdots & \mathbf{D}^{-\left(  l-1\right)
/l}%
\end{array}
\right]  \right\rfloor .
\end{align*}
After carrying out this last operation, the \textquotedblleft
Z\textquotedblright\ matrix expands as follows%
\[
\left[
\begin{array}
[c]{cccc}%
\left\lfloor Z\left(  D^{1/l}\right)  \right\rfloor  & \left\lfloor Z\left(
D^{1/l}\right)  \mathbf{D}^{-1/l}\right\rfloor  & \cdots & \left\lfloor
Z\left(  D^{1/l}\right)  \mathbf{D}^{-\left(  l-1\right)  /l}\right\rfloor \\
\left\lfloor \mathbf{D}^{1/l}Z\left(  D^{1/l}\right)  \right\rfloor  &
\left\lfloor \mathbf{D}^{1/l}Z\left(  D^{1/l}\right)  \mathbf{D}%
^{-1/l}\right\rfloor  & \cdots & \left\lfloor \mathbf{D}^{1/l}Z\left(
D^{1/l}\right)  \mathbf{D}^{-\left(  l-1\right)  /l}\right\rfloor \\
\vdots & \vdots & \ddots & \vdots\\
\left\lfloor \mathbf{D}^{\left(  l-1\right)  /l}Z\left(  D^{1/l}\right)
\right\rfloor  & \left\lfloor \mathbf{D}^{\left(  l-1\right)  /l}Z\left(
D^{1/l}\right)  \mathbf{D}^{-1/l}\right\rfloor  & \cdots & \left\lfloor
\mathbf{D}^{\left(  l-1\right)  /l}Z\left(  D^{1/l}\right)  \mathbf{D}%
^{-\left(  l-1\right)  /l}\right\rfloor
\end{array}
\right]  ,
\]
and the \textquotedblleft X\textquotedblright\ matrix expands similarly. The
technique has the desired effect of expanding each matrix by a\ factor of $l$
because the matrix is $l$ times the original size and because the fractional
powers and flooring operations pick out the terms that should appear in the
$l$-expanded matrix.

\section{Polynomial Symplectic Gram-Schmidt Procedure}

\label{sec:GS}In general, a given set of generators may have complicated
commutation relations. We have to simplify the commutation relations so that
we can encode information qubits with the help of ancilla qubits and halves of
ebits shared with the receiver. In this section, we begin with an arbitrary
set of convolutional generators. We show how to determine a set of generators
with equivalent error-correcting properties and commutation relations that are
the same as those of halves of ebits and ancilla qubits. We first show the
technique for an example by illustrating it for Pauli sequences, for the
quantum check matrix, and with the shifted symplectic product matrix. We then
state a polynomial symplectic Gram-Schmidt orthogonalization algorithm that
performs this action for an arbitrary set of quantum convolutional generators.

\subsection{Example of the Polynomial Symplectic Gram-Schmidt
Orthogonalization Procedure}

\subsubsection{Pauli Picture}

Let us consider again our example from the previous section. Specifically,
consider the respective expressions in (\ref{eq:pauli-conv-simple})\ and
(\ref{eq:two-expanded-code}) for the convolutional generator and the block
code. Recall that we can multiply the generators in a block code without
changing the error-correcting properties of the code \cite{book2000mikeandike}%
. Therefore, we can multiply the sixth generator in
(\ref{eq:two-expanded-code}) to the fourth. We can also multiply the sixth and
the fourth to the second to yield the following equivalent code:%
\begin{equation}
\left.
\begin{array}
[c]{cc}%
X & Z\\
I & X\\
I & I\\
I & I\\
I & I\\
I & I
\end{array}
\right\vert
\begin{array}
[c]{cc}%
I & I\\
Z & X\\
X & Z\\
I & X\\
I & I\\
I & I
\end{array}
\left\vert
\begin{array}
[c]{cc}%
I & I\\
Z & X\\
I & I\\
Z & X\\
X & Z\\
I & X
\end{array}
\right.  . \label{eq:equiv-set}%
\end{equation}
We have manipulated the two-expanded matrix rather than the code in
(\ref{eq:pauli-block-simple})\ because the commutation relations of the above
code are equivalent to the commutation relations of the following operators:%
\[
\left.
\begin{array}
[c]{cc}%
I & Z\\
I & X\\
I & I\\
I & I\\
I & I\\
I & I
\end{array}
\right\vert
\begin{array}
[c]{cc}%
I & I\\
I & I\\
I & Z\\
I & X\\
I & I\\
I & I
\end{array}
\left\vert
\begin{array}
[c]{cc}%
I & I\\
I & I\\
I & I\\
I & I\\
I & Z\\
I & X
\end{array}
\right.  .
\]
We can use three ebits to encode the set of generators in (\ref{eq:equiv-set})
because they have the same commutation relations as the above operators and
the above operators correspond to halves of three ebits. We resolve the
anticommutation relations by using the following entanglement-assisted code:%
\[
\left.
\begin{array}
[c]{ccc}%
Z & X & Z\\
X & I & X\\
I & I & I\\
I & I & I\\
I & I & I\\
I & I & I
\end{array}
\right\vert
\begin{array}
[c]{ccc}%
I & I & I\\
I & Z & X\\
Z & X & Z\\
X & I & X\\
I & I & I\\
I & I & I
\end{array}
\left\vert
\begin{array}
[c]{ccc}%
I & I & I\\
I & Z & X\\
I & I & I\\
I & Z & X\\
Z & X & Z\\
X & I & X
\end{array}
\right.  .
\]
The convention above is that the first qubit of each frame belongs to Bob and
corresponds to half of an ebit. The second two qubits of each frame belong to
Alice. The overall code forms a commuting stabilizer so that it corresponds to
a valid quantum code. Bob could measure the above operators to diagnose errors
or he could measure the following operators that are equivalent by row
operations:%
\[
\left.
\begin{array}
[c]{ccc}%
Z & X & Z\\
X & I & X\\
I & I & I\\
I & I & I\\
I & I & I\\
I & I & I
\end{array}
\right\vert
\begin{array}
[c]{ccc}%
I & I & I\\
X & Z & I\\
Z & X & Z\\
X & I & X\\
I & I & I\\
I & I & I
\end{array}
\left\vert
\begin{array}
[c]{ccc}%
I & I & I\\
I & I & I\\
I & I & I\\
X & Z & I\\
Z & X & Z\\
X & I & X
\end{array}
\right.  .
\]
One can check that the operators corresponding to the second two qubits of
each frame are equivalent to the desired generators in
(\ref{eq:two-expanded-code}).

\subsubsection{Polynomial Picture}

Let us consider the convolutional generator in (\ref{eq:pauli-conv-simple}).
We now use the polynomial formalism because it is easier to perform the
manipulations for general codes in this picture rather than in the Pauli picture.

We extend the row operations from the above block code to the polynomial
picture. Each row operation multiplied the even-numbered generators by
themselves shifted by two qubits. Extending this operation to act on an
infinite Pauli sequence corresponds to multiplying the second generator in
(\ref{eq:two-exp-check})\ by the rational polynomial $1/\left(  1+D\right)  $.
Consider the two-expanded check matrix with the operation described above
applied to the second generator:%
\[
\left[
\begin{array}
[c]{c}%
g_{1}\left(  D\right) \\
g_{2}\left(  D\right)
\end{array}
\right]  =\left[  \left.
\begin{array}
[c]{cc}%
0 & 1\\
\frac{D}{1+D} & 0
\end{array}
\right\vert
\begin{array}
[c]{cc}%
1 & 0\\
0 & \frac{1}{1+D}%
\end{array}
\right]  .
\]
The shifted symplectic products $\left(  g_{1}\odot g_{1}\right)  \left(
D\right)  =0$, $\left(  g_{1}\odot g_{2}\right)  \left(  D\right)  =1$,
$\ $and $\left(  g_{2}\odot g_{2}\right)  \left(  D\right)  =0$ capture the
commutation relations of the resulting code for all frames. We can therefore
resolve this anticommutativity by prepending a column that corresponds to Bob
possessing an ebit:%
\[
\left[  \left.
\begin{array}
[c]{ccc}%
1 & 0 & 1\\
0 & \frac{D}{1+D} & 0
\end{array}
\right\vert
\begin{array}
[c]{ccc}%
0 & 1 & 0\\
1 & 0 & \frac{1}{1+D}%
\end{array}
\right]  .
\]
Bob possesses the qubit corresponding to column one of both the
\textquotedblleft Z\textquotedblright\ and \textquotedblleft
X\textquotedblright\ matrix and Alice possesses the two qubits corresponding
to the second and third columns. The code above has equivalent
error-correcting properties to those of the desired generators.

The generators in the above polynomial setting correspond to Pauli sequences
with infinite weight. Infinite-weight generators are undesirable because Bob
cannot measure an infinite number of qubits. There is a simple solution to
this problem and it is similar to what we did at the end of the previous
subsection. We multiply the second row of the above check matrix by $1+D$ to
obtain the following check matrix that has equivalent error-correcting
properties to the above one:%
\[
\left[  \left.
\begin{array}
[c]{ccc}%
1 & 0 & 1\\
0 & D & 0
\end{array}
\right\vert
\begin{array}
[c]{ccc}%
0 & 1 & 0\\
1+D & 0 & 1
\end{array}
\right]  .
\]
The generators in the above check matrix correspond to Pauli sequences with
finite weight by the P2B isomorphism. Specifically, the above two generators
are equivalent to the following two Pauli sequences and all of their
three-qubit shifts:%
\[
\cdots\left\vert
\begin{array}
[c]{ccc}%
I & I & I\\
I & I & I
\end{array}
\right\vert
\begin{array}
[c]{ccc}%
Z & X & Z\\
X & I & X
\end{array}
\left\vert
\begin{array}
[c]{ccc}%
I & I & I\\
X & Z & I
\end{array}
\right\vert \left.
\begin{array}
[c]{ccc}%
I & I & I\\
I & I & I
\end{array}
\right\vert \cdots
\]
Bob can measure these operators because they have finite weight.

\subsubsection{Shifted Symplectic Product Matrix Picture}

We can more easily determine the row operations that simplify the commutation
relations by looking only at the shifted symplectic product matrix. We would
like to perform row operations on the check matrix to reduce its corresponding
shifted symplectic product matrix to the standard form in
(\ref{eq:standard-symp-form}), so that it has commutation relations equivalent
to those of halves of ebits and ancilla qubits.

The shifted symplectic product matrix corresponding to the check matrix in
(\ref{eq:poly-conv-simple}) is a one-element matrix $\Omega_{g}\left(
D\right)  $ where%
\[
\Omega_{g}\left(  D\right)  =\left[  D^{-1}+D\right]  .
\]
It is clear that $\Omega_{g}\left(  D\right)  $ is not reducible by row
operations to the $2\times2$ matrix $J$ in (\ref{eq:J-matrix})\ because
$\Omega_{g}\left(  D\right)  $ is a one-element matrix. It is also not
reducible to the one-element null matrix $\left[  0\right]  $ because there is
no valid row operation that can zero out the term $D^{-1}+D$.

We therefore consider the two-expanded check matrix in (\ref{eq:two-exp-check}%
) to determine if we can reduce it with row operations to the standard form in
(\ref{eq:standard-symp-form}). The shifted symplectic product matrix
$\Omega_{G_{2}}\left(  D\right)  $ for the two-expanded check matrix
$G_{2}\left(  D\right)  $ is as follows:%
\[
\Omega_{G_{2}}\left(  D\right)  =\left[
\begin{array}
[c]{cc}%
0 & 1+D^{-1}\\
1+D & 0
\end{array}
\right]  .
\]
We can represent the row operation of multiplying the second generator by
$1/\left(  1+D\right)  $ as a matrix $R\left(  D\right)  $ where%
\[
R\left(  D\right)  =\left[
\begin{array}
[c]{cc}%
1 & 0\\
0 & 1/\left(  1+D\right)
\end{array}
\right]  .
\]
The effect on the matrix $\Omega_{G_{2}}\left(  D\right)  $ is to change it to%
\[
R\left(  D\right)  \Omega_{G_{2}}\left(  D\right)  R^{T}\left(  D^{-1}\right)
=J,
\]
as described in (\ref{eq:symp-row-op}). The above matrix $J$ has equivalent
commutation relations to half of an ebit. We can therefore use one ebit per
two qubits to encode this code.

In a later section, we show how to devise encoding and decoding circuits,
beginning from $k$ information qubits, $c$ ebits, and $a$ ancilla qubits per
frame. Before that, however, we present a procedure to reduce the commutation
relations of any check matrix to those of ebits and ancilla qubits.

\subsection{The Polynomial Symplectic Gram-Schmidt Orthogonalization Procedure
for General Codes}

We detail a polynomial version of the symplectic Gram-Schmidt
orthogonalization procedure in this section. It is a generalization of the
procedure we developed for the above example. Before detailing the algorithm,
we first prove a lemma that shows how to determine the shifted symplectic
product matrix for an $l$-expanded version of the generators by starting from
the shifted symplectic product matrix of the original generators.

\subsubsection{The Shifted Symplectic Product Matrix for an $l$-Expanded Code}

Suppose the shifted symplectic product matrix $\Omega\left(  D\right)  $ of a
given check matrix $H\left(  D\right)  $ is as follows:%
\[
\Omega\left(  D\right)  =Z\left(  D\right)  X^{T}\left(  D^{-1}\right)
+X\left(  D\right)  Z^{T}\left(  D^{-1}\right)  .
\]

\begin{lemma}
The shifted symplectic product matrix $\Omega_{l}\left(  D\right)  $ of the
$l$-expanded check matrix $H_{l}\left(  D\right)  $ is as follows:%
\[
\Omega_{l}\left(  D\right)  =\left\lfloor R_{l}\left(  D^{1/l}\right)
\Omega\left(  D^{1/l}\right)  R_{l}^{T}\left(  D^{-1/l}\right)  \right\rfloor
,
\]
where the flooring operation $\left\lfloor \cdot\right\rfloor $ nulls the
coefficients of any fractional power of \thinspace$D$.
\end{lemma}

\begin{proof}
Consider that the \textquotedblleft X\textquotedblright\ matrix $X_{l}\left(
D\right)  $ of the $l$-expanded check matrix $H_{l}\left(  D\right)  $ is as
follows:%
\[
X_{l}\left(  D\right)  =\left\lfloor R_{l}\left(  D^{1/l}\right)  X\left(
D^{1/l}\right)  C_{l}^{\prime}\left(  D^{1/l}\right)  \right\rfloor ,
\]
where
\[
C_{l}^{\prime}\left(  D^{1/l}\right)  \equiv\left[
\begin{array}
[c]{cccc}%
\mathbf{D}^{0} & \mathbf{D}^{-1} & \cdots & \mathbf{D}^{-\left(  l-1\right)  }%
\end{array}
\right]  ,
\]
and each diagonal $\mathbf{D}$ matrix is $n\times n$-dimensional. It is also
then true that the \textquotedblleft Z\textquotedblright\ matrix of
$H_{l}\left(  D\right)  $ is as follows:%
\[
Z_{l}\left(  D\right)  =\left\lfloor R_{l}\left(  D^{1/l}\right)  Z\left(
D^{1/l}\right)  C_{l}^{\prime}\left(  D^{1/l}\right)  \right\rfloor .
\]
The matrix transpose operation, the time reversal operation (substituting
$D^{-1}$ for $D$), and matrix addition are all invariant under the flooring
operation $\left\lfloor \cdot\right\rfloor $ for arbitrary matrices $M\left(
D\right)  $ and $N\left(  D\right)  $:%
\begin{align*}
\left\lfloor M^{T}\left(  D^{1/l}\right)  \right\rfloor  &  =\left\lfloor
M\left(  D^{1/l}\right)  \right\rfloor ^{T},\\
\left\lfloor M\left(  D^{-1/l}\right)  \right\rfloor  &  =\left.  \left\lfloor
M\left(  D^{1/l}\right)  \right\rfloor \right\vert _{D=D^{-1}},\\
\left\lfloor M\left(  D^{1/l}\right)  +N\left(  D^{1/l}\right)  \right\rfloor
&  =\left\lfloor M\left(  D^{1/l}\right)  \right\rfloor +\left\lfloor N\left(
D^{1/l}\right)  \right\rfloor .
\end{align*}
Additionally, the following property holds for two arbitrary binary
polynomials $f\left(  D\right)  $ and $g\left(  D\right)  $:%
\begin{equation}
\left\lfloor f\left(  D^{1/l}\right)  g\left(  D^{1/l}\right)  \right\rfloor
=\sum_{i=0}^{l-1}\left\lfloor D^{-i/l}f\left(  D^{1/l}\right)  \right\rfloor
\left\lfloor D^{i/l}g\left(  D^{1/l}\right)  \right\rfloor
.\label{eq:mult-prop-floor}%
\end{equation}
Now consider the product $X_{l}\left(  D\right)  Z_{l}^{T}\left(
D^{-1}\right)  $:%
\begin{align*}
X_{l}\left(  D\right)  Z_{l}^{T}\left(  D^{-1}\right)   &  =\left\lfloor
R_{l}\left(  D^{1/l}\right)  X\left(  D^{1/l}\right)  C_{l}^{\prime}\left(
D^{1/l}\right)  \right\rfloor \ \left.  \left(  \left\lfloor R_{l}\left(
D^{1/l}\right)  Z\left(  D^{1/l}\right)  C_{l}^{\prime}\left(  D^{1/l}\right)
\right\rfloor \right)  ^{T}\right\vert _{D=D^{-1}}\\
&  =\left\lfloor R_{l}\left(  D^{1/l}\right)  X\left(  D^{1/l}\right)
C_{l}^{\prime}\left(  D^{1/l}\right)  \right\rfloor \ \left\lfloor
C_{l}^{\prime T}\left(  D^{-1/l}\right)  Z^{T}\left(  D^{-1/l}\right)
R_{l}^{T}\left(  D^{-1/l}\right)  \right\rfloor \\
&  =\sum_{i=0}^{l-1}\left\lfloor D^{-i/l}R_{l}\left(  D^{1/l}\right)  X\left(
D^{1/l}\right)  \right\rfloor \ \left\lfloor D^{i/l}Z^{T}\left(
D^{-1/l}\right)  R_{l}^{T}\left(  D^{-1/l}\right)  \right\rfloor \\
&  =\left\lfloor R_{l}\left(  D^{1/l}\right)  X\left(  D^{1/l}\right)
Z^{T}\left(  D^{-1/l}\right)  R_{l}^{T}\left(  D^{-1/l}\right)  \right\rfloor
,
\end{align*}
where the second line uses the invariance of the flooring operation with
respect to matrix transposition and time reversal, the third line expands the
matrix multiplications using the matrix $C_{l}^{\prime}\left(  D\right)  $
defined above, and the last line uses the matrix generalization of the
multiplication property defined in (\ref{eq:mult-prop-floor}). Our final step
is to use the invariance of the flooring operation with respect to matrix
addition:%
\begin{align*}
\Omega_{l}\left(  D\right)   &  =X_{l}\left(  D\right)  Z_{l}^{T}\left(
D^{-1}\right)  +Z_{l}\left(  D\right)  X_{l}^{T}\left(  D^{-1}\right)  ,\\
&  =\left\lfloor R_{l}\left(  D^{1/l}\right)  X\left(  D^{1/l}\right)
Z^{T}\left(  D^{-1/l}\right)  R_{l}^{T}\left(  D^{-1/l}\right)  \right\rfloor
+\left\lfloor R_{l}\left(  D^{1/l}\right)  Z\left(  D^{1/l}\right)
X^{T}\left(  D^{-1/l}\right)  R_{l}^{T}\left(  D^{-1/l}\right)  \right\rfloor
\\
&  =\left\lfloor
\begin{array}
[c]{c}%
R_{l}\left(  D^{1/l}\right)  X\left(  D^{1/l}\right)  Z^{T}\left(
D^{-1/l}\right)  R_{l}^{T}\left(  D^{-1/l}\right)  \\
+R_{l}\left(  D^{1/l}\right)  Z\left(  D^{1/l}\right)  X^{T}\left(
D^{-1/l}\right)  R_{l}^{T}\left(  D^{-1/l}\right)
\end{array}
\right\rfloor \\
&  =\left\lfloor R_{l}\left(  D^{1/l}\right)  \Omega\left(  D^{1/l}\right)
R_{l}^{T}\left(  D^{-1/l}\right)  \right\rfloor .
\end{align*}
\end{proof}

\subsubsection{The Gram-Schmidt Procedure}

We now present the polynomial symplectic Gram-Schmidt orthogonalization
procedure that reduces the commutation relations of a given set of
convolutional generators to have the standard form in
(\ref{eq:standard-symp-form}).

Consider the following $n-k\times2n$-dimensional quantum check matrix
$H\left(  D\right)  $:%
\[
H\left(  D\right)  =\left[  \left.
\begin{array}
[c]{c}%
Z\left(  D\right)
\end{array}
\right\vert
\begin{array}
[c]{c}%
X\left(  D\right)
\end{array}
\right]  .
\]
Label each row as $h_{i}\left(  D\right)  =\left[  \left.
\begin{array}
[c]{c}%
z_{i}\left(  D\right)
\end{array}
\right\vert
\begin{array}
[c]{c}%
x_{i}\left(  D\right)
\end{array}
\right]  $ for all $i\in\left\{  1,\ldots,n-k\right\}  $.

We state the Gram-Schmidt procedure in terms of its effect on the shifted
symplectic product matrix. It is easier to see how the algorithm proceeds by
observing the shifted symplectic product matrix rather than by tracking the
generators in the check matrix.

Let $l$ denote the amount by which we expand the check matrix $H\left(
D\right)  $. Suppose first that $l=1$ (we do not expand the check matrix). Let
us say that the check matrix has $r$ generators (we take a number different
from $n-k$ because the number of generators may not be equal to $n-k$ for
future iterations). There are the following possibilities:

\begin{enumerate}
\item There is a generator $h_{i}\left(  D\right)  $ such that $\left(
h_{i}\odot h_{j}\right)  \left(  D\right)  =0$ for all $j\in\left\{
1,\ldots,r\right\}  $. In this case, the generator is already decoupled from
all the others and corresponds to an ancilla qubit because an ancilla and it
share the same commutation relations. Swap $h_{i}\left(  D\right)  $ to be the
first row of the matrix so that it is $h_{1}\left(  D\right)  $. The shifted
symplectic product matrix then has the form:%
\[%
\begin{bmatrix}
0 & 0 & \cdots & 0\\
0 & h_{2,2} & \cdots & h_{2,r}\\
\vdots & \vdots & \ddots & \vdots\\
0 & h_{r,2} & \cdots & h_{r,r}%
\end{bmatrix}
=%
\begin{bmatrix}
0
\end{bmatrix}
\oplus%
\begin{bmatrix}
h_{2,2} & \cdots & h_{2,r}\\
\vdots & \ddots & \vdots\\
h_{r,2} & \cdots & h_{r,r}%
\end{bmatrix}
,
\]
where $\left[  0\right]  $ is the one-element zero matrix and we use the
shorthand $h_{i,j}=\left(  h_{i}\odot h_{j}\right)  \left(  D\right)  $. We
remove generator $h_{1}\left(  D\right)  $ from check matrix $H\left(
D\right)  $ and continue to step two below for the remaining generators in the matrix.

\item There are two generators $h_{i}\left(  D\right)  $ and $h_{j}\left(
D\right)  $ for which $\left(  h_{i}\odot h_{j}\right)  \left(  D\right)
=D^{m}$ for some integer $m$ and $\left(  h_{i}\odot h_{i}\right)  \left(
D\right)  =\left(  h_{j}\odot h_{j}\right)  \left(  D\right)  =0$. In this
case these generators correspond exactly to an ebit. Multiply generator
$h_{j}\left(  D\right)  $ by $D^{m}$. This row operation has the effect of
delaying (or advancing) the generator by an amount $m$ and changes the shifted
symplectic product to be $\left(  h_{i}\odot h_{j}\right)  \left(  D\right)
=1$. These two generators considered by themselves now have the commutation
relations of half of an ebit. Swap the generators $h_{i}\left(  D\right)  $
and $h_{j}\left(  D\right)  $ to be the first and second respective rows of
the check matrix $H\left(  D\right)  $. Call them $h_{1}\left(  D\right)  $
and $h_{2}\left(  D\right)  $ respectively. The shifted symplectic product
matrix is then as follows:%
\[%
\begin{bmatrix}
0 & 1 & h_{1,3} & \cdots & h_{1,r}\\
1 & 0 & h_{2,3} & \cdots & h_{2,r}\\
h_{3,1} & h_{3,2} & h_{3,3} & \cdots & h_{3,r}\\
\vdots & \vdots & \vdots & \ddots & \vdots\\
h_{r,2} & h_{r,2} & h_{r,3} & \cdots & h_{r,r}%
\end{bmatrix}
.
\]
We use the following row operations to decouple the other generators from
these two generators:%
\[
h_{i}^{\prime}\left(  D\right)  \equiv h_{i}\left(  D\right)  +\left(
h_{i}\odot h_{2}\right)  \left(  D\right)  \cdot h_{1}\left(  D\right)
+\left(  h_{i}\odot h_{1}\right)  \left(  D\right)  \cdot h_{2}\left(
D\right)  \text{ for all }i\in\left\{  3,\ldots,r\right\}  .
\]
The shifted symplectic product matrix becomes as follows under these row
operations:%
\[%
\begin{bmatrix}
0 & 1 & 0 & \cdots & 0\\
1 & 0 & 0 & \cdots & 0\\
0 & 0 & h_{3,3}^{\prime} & \cdots & h_{3,r}^{\prime}\\
\vdots & \vdots & \vdots & \ddots & \vdots\\
0 & 0 & h_{r,3}^{\prime} & \cdots & h_{r,r}^{\prime}%
\end{bmatrix}
=J\oplus%
\begin{bmatrix}
h_{3,3}^{\prime} & \cdots & h_{3,r}^{\prime}\\
\vdots & \ddots & \vdots\\
h_{r,3}^{\prime} & \cdots & h_{r,r}^{\prime}%
\end{bmatrix}
.
\]
The first two generators are now decoupled from the other generators and have
the commutation relations of half of an ebit. We remove the first two
generators from the check matrix so that it now consists of generators
$h_{3}^{\prime}\left(  D\right)  $, $\ldots$, $h_{r}^{\prime}\left(  D\right)
$. We check to see if the conditions at the beginning of the previous step or
this step hold for any other generators. If so, repeat the previous step or
this step on the remaining generators. If not, see if the conditions for step
three hold.

\item There are two generators $h_{i}\left(  D\right)  $ and $h_{j}\left(
D\right)  $ for which $\left(  h_{i}\odot h_{i}\right)  \left(  D\right)
=\left(  h_{j}\odot h_{j}\right)  \left(  D\right)  =0$ but $\left(
h_{i}\odot h_{j}\right)  \left(  D\right)  \neq D^{m}$ for some $m$ and
$\left(  h_{i}\odot h_{j}\right)  \left(  D\right)  \neq0$. Multiply generator
$h_{j}\left(  D\right)  $ by $1/\left(  h_{j}\odot h_{i}\right)  \left(
D\right)  $. Generator $h_{j}\left(  D\right)  $ becomes infinite weight
because $\left(  h_{j}\odot h_{i}\right)  \left(  D\right)  $ is a polynomial
with two or more powers of $D$ with non-zero coefficients. Now the shifted
symplectic product relations are as follows: $\left(  h_{i}\odot h_{i}\right)
\left(  D\right)  =\left(  h_{j}\odot h_{j}\right)  \left(  D\right)  =0$ and
$\left(  h_{i}\odot h_{j}\right)  \left(  D\right)  =1$. We handle this case
as we did the previous case after the two generators there had the commutation
relations of half of an ebit.

\item None of these conditions hold. In this case, we stop this iteration of
the algorithm and expand the check matrix by the next factor $l:=l+1$ and
repeat the above steps.
\end{enumerate}

We have not proven that this procedure converges on all codes; however, it
does converge on all the codes we have tried. We conjecture that this
procedure converges for all codes, but even if this is true, in principle it
might require expansion to a large number of generators. A simple and
practical convergence condition is as follows. In practice, convolutional
codes do not act on an infinite stream of qubits but instead act on a finite
number of qubits. It may be that there are codes for which this procedure
either does not converge or must be expanded until the frame size of the
expanded code exceeds the number of qubits that the code acts on. In this
case, we would not employ this procedure, and instead would treat the code as
a block code, where we could employ the methods from
Refs.~\cite{science2006brun,arx2006brun} for encoding and decoding. It is
unclear if this practical convergence condition will ever really be necessary.
This procedure does converge for a large number of useful codes, so that the
frame size of the expanded code is much less than the number of qubits that
the code acts on, and we have not found an example where this procedure fails.

\section{Encoding and Decoding Entanglement-Assisted Quantum Convolutional
Codes}

\label{sec:encode-decode}This section proves the main theorem of this paper.
The theorem assumes that we have already processed an arbitrary check matrix
with the Gram-Schmidt algorithm and that the shifted symplectic product matrix
corresponding to the processed check matrix has the standard form in
(\ref{eq:standard-symp-form}). The theorem shows how to encode a set of
information qubits, ancilla qubits, and halves of ebits into a code that has
equivalent error-correcting properties to those of a desired set of
convolutional generators.

The theorem uses both finite-depth and infinite-depth operations in the
encoding circuit and finite-depth operations in the decoding circuit. The
finite-depth property of the operations in the decoding circuit guarantees
that catastrophic error propagation does not occur when decoding the encoded
qubit stream. We do not review finite-depth or infinite-depth operations here
but instead refer the reader to our previous article \cite{arx2007wildeEAQCC}%
\ that has a detailed discussion of these operations.

\begin{theorem}
\label{thm:main}Suppose we have a set of quantum convolutional generators in
the $n-k\times2n$-dimensional matrix $H\left(  D\right)  $ where%
\[
H\left(  D\right)  =\left[  \left.
\begin{array}
[c]{c}%
Z\left(  D\right)
\end{array}
\right\vert
\begin{array}
[c]{c}%
X\left(  D\right)
\end{array}
\right]  .
\]
Its shifted symplectic product matrix $\Omega\left(  D\right)  $ is as
follows:%
\[
\Omega\left(  D\right)  =Z\left(  D\right)  X^{T}\left(  D^{-1}\right)
+X\left(  D\right)  Z^{T}\left(  D^{-1}\right)  .
\]
Suppose the check matrix $H\left(  D\right)  $ is the matrix resulting from
processing with the polynomial symplectic Gram-Schmidt orthogonalization
procedure. Therefore, $\Omega\left(  D\right)  $ has the standard form in
(\ref{eq:standard-symp-form}) with parameters $c$ and $a=n-k-2c$. Then there
exists an online encoding circuit for the code that uses finite-depth and
infinite-depth operations in the shift-invariant Clifford group
\cite{arx2007wildeEAQCC} and there exists an online decoding circuit for the
code that uses finite-depth operations. The code encodes $k+c$ information
qubits per frame with the help of $c$ ebits and $a=n-k-2c$ ancilla qubits.
\end{theorem}

\begin{proof}
We prove the theorem by giving an algorithm to compute both the encoding
circuit and the decoding circuit. The shifted symplectic product matrix
$\Omega\left(  D\right)  $ for check matrix $H\left(  D\right)  $ is in
standard form so that the first $2c$ rows have the commutation relations of
$c$ halves of ebits and the last $a$ rows have the commutation relations of
$a$ ancilla qubits. Perform the algorithm outlined in
Refs.~\cite{ieee2006grassl,isit2006grassl}\ on the last $a$ generators that
correspond to the ancilla qubits. The algorithm uses finite-depth CNOT\ gates,
Hadamard gates, and phase gates. The resulting check matrix has the following
form:%
\begin{equation}
\left[  \left.
\begin{array}
[c]{cc}%
Z^{\prime\prime}\left(  D\right)  & Z^{\prime}\left(  D\right) \\
\Gamma\left(  D\right)  & 0
\end{array}
\right\vert
\begin{array}
[c]{cc}%
0 & X^{\prime}\left(  D\right) \\
0 & 0
\end{array}
\right]  , \label{eq:start-finite-depth-encode}%
\end{equation}
where the matrix $Z^{\prime\prime}\left(  D\right)  $ and the null matrix at
the top left of the \textquotedblleft X\textquotedblright\ matrix each have
dimension $2c\times a$, the matrices $Z^{\prime}\left(  D\right)  $ and
$X^{\prime}\left(  D\right)  $ each have dimension $2c\times n-a$, and all
matrices in the second set of rows each have $a$ rows. The matrix
$\Gamma\left(  D\right)  $ may have entries that are rational polynomials. If
so, replace each of these entries with a \textquotedblleft1\textquotedblright%
\ so that the resulting matrix has the following form:%
\[
\left[  \left.
\begin{array}
[c]{cc}%
Z^{\prime\prime}\left(  D\right)  & Z^{\prime}\left(  D\right) \\
I & 0
\end{array}
\right\vert
\begin{array}
[c]{cc}%
0 & X^{\prime}\left(  D\right) \\
0 & 0
\end{array}
\right]  .
\]
This replacement is equivalent to taking a subcode of the original that has
equivalent error-correcting properties and rate~\cite{isit2006grassl}. We can
also think of the replacement merely as row operations with rational
polynomials. We then perform row operations from the last $a$ rows to the
first $2c$ rows to obtain the following check matrix:%
\[
\left[  \left.
\begin{array}
[c]{cc}%
0 & Z^{\prime}\left(  D\right) \\
I & 0
\end{array}
\right\vert
\begin{array}
[c]{cc}%
0 & X^{\prime}\left(  D\right) \\
0 & 0
\end{array}
\right]  .
\]
The shifted symplectic product matrix still has the standard form in
(\ref{eq:standard-symp-form}) because these last row operations do not change
its entries.
We now focus exclusively on the first $2c$ rows because the previous steps
decoupled the last $a$ rows from the first $2c$ rows. Consider the following
submatrix:%
\[
H^{\prime}\left(  D\right)  =\left[  \left.
\begin{array}
[c]{c}%
Z\left(  D\right)
\end{array}
\right\vert
\begin{array}
[c]{c}%
X\left(  D\right)
\end{array}
\right]  ,
\]
where we have reset variable labels so that $Z\left(  D\right)  =Z^{\prime
}\left(  D\right)  $ and $X\left(  D\right)  =X^{\prime}\left(  D\right)  $.
Perform row permutations on the above matrix so that the shifted symplectic
product matrix for $H^{\prime}\left(  D\right)  $ changes from%
\[%
%TCIMACRO{\dbigoplus \limits_{i=1}^{c}}%
%BeginExpansion
{\displaystyle\bigoplus\limits_{i=1}^{c}}
%EndExpansion
J,
\]
to become%
\begin{equation}
\left[
\begin{array}
[c]{cc}%
0 & I\\
I & 0
\end{array}
\right]  , \label{eq:alt-symp-rels}%
\end{equation}
where each identity and null matrix in the above matrix is $c\times c$-dimensional.
We can employ the algorithm from Ref.~\cite{ieee2006grassl} on the first $c$
generators because the first $c$ rows of the resulting check matrix form a
commuting set (we do not use the row operations in that algorithm). The
algorithm employs finite-depth CNOT\ gates, Hadamard gates, and phase gates
and reduces the check matrix to have the following form:%
\[
\left[  \left.
\begin{array}
[c]{cc}%
0 & 0\\
U\left(  D\right)  & Z_{2}\left(  D\right)
\end{array}
\right\vert
\begin{array}
[c]{cc}%
L\left(  D\right)  & 0\\
X_{1}\left(  D\right)  & X_{2}\left(  D\right)
\end{array}
\right]  ,
\]
where $L\left(  D\right)  $ is a $c\times c$ lower triangular matrix and
$U\left(  D\right)  $ is a $c\times c$ upper triangular matrix. The
$i^{\text{th}}$ diagonal entry $u_{ii}\left(  D\right)  $ of $U\left(
D\right)  $ is equal to $1/l_{ii}\left(  D^{-1}\right)  $ where $l_{ii}\left(
D\right)  $ is the $i^{\text{th}}$ diagonal entry of $L\left(  D\right)  $.
This relationship holds because of the shifted symplectic relations in
(\ref{eq:alt-symp-rels}) and because gates in the shift-invariant Clifford
group do not affect the shifted symplectic relations.
We now employ several row operations whose net effect is to preserve the
shifted symplectic relations in (\ref{eq:alt-symp-rels})---we can therefore
include them as a part of the original polynomial symplectic Gram-Schmidt
orthogonalization procedure. Multiply row $i$ of the above check matrix by
$1/l_{ii}\left(  D\right)  $ and multiply row $i+c$ by $1/u_{ii}\left(
D\right)  $ for all $i\in\left\{  1,\ldots,c\right\}  $. Then use row
operations to cancel all the off-diagonal entries in both $L\left(  D\right)
$ and $U\left(  D\right)  $. The resulting check matrix has the following
form:%
\[
\left[  \left.
\begin{array}
[c]{cc}%
0 & 0\\
I & Z_{2}^{\prime}\left(  D\right)
\end{array}
\right\vert
\begin{array}
[c]{cc}%
I & 0\\
X_{1}^{\prime}\left(  D\right)  & X_{2}^{\prime}\left(  D\right)
\end{array}
\right]  ,
\]
where the primed matrices result from all the row operations. One can check
that the shifted symplectic relations of the above matrix are equivalent to
those in (\ref{eq:alt-symp-rels}). Perform Hadamard gates on the first $c$
qubits. The check matrix becomes%
\begin{equation}
\left[  \left.
\begin{array}
[c]{cc}%
I & 0\\
X_{1}^{\prime}\left(  D\right)  & Z_{2}^{\prime}\left(  D\right)
\end{array}
\right\vert
\begin{array}
[c]{cc}%
0 & 0\\
I & X_{2}^{\prime}\left(  D\right)
\end{array}
\right]  . \label{eq:desired-check-matrix}%
\end{equation}
We show how to encode the above matrix starting from $c$ ebits and $k+c$
information qubits. The following matrix stabilizes a set of $c$ ebits:%
\begin{equation}
\left[  \left.
\begin{array}
[c]{ccc}%
I & I & 0\\
0 & 0 & 0
\end{array}
\right\vert
\begin{array}
[c]{ccc}%
0 & 0 & 0\\
I & I & 0
\end{array}
\right]  ,\label{eq:start-encode}%
\end{equation}
where each identity matrix is $c\times c$ and the last column of zeros in each
matrix is $c\times\left(  k+c\right)  $. The receiver Bob possesses the first
$c$ qubits and the sender Alice possesses the last $k+2c$ qubits. The
following matrix is the \textquotedblleft information-qubit\textquotedblright%
\ matrix \cite{arx2007wildeEAQCC}:%
\[
\left[  \left.
\begin{array}
[c]{ccc}%
0 & 0 & I\\
0 & 0 & 0
\end{array}
\right\vert
\begin{array}
[c]{ccc}%
0 & 0 & 0\\
0 & 0 & I
\end{array}
\right]  ,
\]
where each identity matrix is $\left(  k+c\right)  \times\left(  k+c\right)  $
and each column of zeros is $\left(  k+c\right)  \times c$. It is important to
track the information-qubit matrix throughout encoding and decoding so that we
can determine at the end of the process if we have truly decoded the
information qubits. Perform finite-depth CNOT\ operations from the first $c$
ebits to the last $k+c$ qubits to encode the numerators of the entries in
matrix $Z_{2}^{\prime}\left(  D\right)  $. Let $Z_{2,N}^{\prime}\left(
D\right)  $ denote this matrix of the numerators of the entries in
$Z_{2}^{\prime}\left(  D\right)  $. The stabilizer matrix becomes%
\[
\left[  \left.
\begin{array}
[c]{ccc}%
I & I & 0\\
0 & 0 & 0
\end{array}
\right\vert
\begin{array}
[c]{ccc}%
0 & 0 & 0\\
I & I & Z_{2,N}^{\prime}\left(  D\right)
\end{array}
\right]  ,
\]
and the information-qubit matrix becomes%
\[
\left[  \left.
\begin{array}
[c]{ccc}%
0 & Z_{2,N}^{\prime\prime}\left(  D\right)   & I\\
0 & 0 & 0
\end{array}
\right\vert
\begin{array}
[c]{ccc}%
0 & 0 & 0\\
0 & 0 & I
\end{array}
\right]  ,
\]
where $Z_{2,N}^{\prime\prime}\left(  D\right)  $ is the matrix that results on
the \textquotedblleft Z\textquotedblright\ side after performing the CNOT
operations corresponding to the entries in $Z_{2,N}^{\prime}\left(  D\right)
$. Perform Hadamard gates on the last $k+c$ qubits. The stabilizer matrix
becomes%
\[
\left[  \left.
\begin{array}
[c]{ccc}%
I & I & 0\\
0 & 0 & Z_{2,N}^{\prime}\left(  D\right)
\end{array}
\right\vert
\begin{array}
[c]{ccc}%
0 & 0 & 0\\
I & I & 0
\end{array}
\right]  ,
\]
and the information-qubit matrix becomes%
\[
\left[  \left.
\begin{array}
[c]{ccc}%
0 & Z_{2,N}^{\prime\prime}\left(  D\right)   & 0\\
0 & 0 & I
\end{array}
\right\vert
\begin{array}
[c]{ccc}%
0 & 0 & I\\
0 & 0 & 0
\end{array}
\right]  .
\]
Let $X_{2,N}^{\prime}\left(  D\right)  $ denote the matrix whose entries are
the numerators of the entries in $X_{2}^{\prime}\left(  D\right)  $. Perform
CNOT\ gates from the first $c$ qubits to the last $k+c$ qubits corresponding
to the entries in $X_{2,N}^{\prime}\left(  D\right)  $. The stabilizer matrix
becomes%
\[
\left[  \left.
\begin{array}
[c]{ccc}%
I & I & 0\\
0 & A\left(  D\right)   & Z_{2,N}^{\prime}\left(  D\right)
\end{array}
\right\vert
\begin{array}
[c]{ccc}%
0 & 0 & 0\\
I & I & X_{2,N}^{\prime}\left(  D\right)
\end{array}
\right]  ,
\]
where $A\left(  D\right)  =Z_{2,N}^{\prime}\left(  D\right)  X_{2,N}%
^{\prime\prime}\left(  D\right)  $ and $X_{2,N}^{\prime\prime}\left(
D\right)  $ is the matrix that results on the \textquotedblleft
Z\textquotedblright\ side after performing the CNOT operations on the
\textquotedblleft X\textquotedblright\ side corresponding to the entries in
$X_{2,N}^{\prime}\left(  D\right)  $. The information-qubit matrix becomes%
\[
\left[  \left.
\begin{array}
[c]{ccc}%
0 & Z_{2,N}^{\prime\prime}\left(  D\right)   & 0\\
0 & X_{2,N}^{\prime\prime}\left(  D\right)   & I
\end{array}
\right\vert
\begin{array}
[c]{ccc}%
0 & 0 & I\\
0 & 0 & 0
\end{array}
\right]  .
\]
Let $\Gamma\left(  D\right)  $ be a diagonal matrix whose $i^{\text{th}}$
diagonal entry is the denominator of the $i^{\text{th}}$ row of $Z_{2}%
^{\prime}\left(  D\right)  $ and $X_{2}^{\prime}\left(  D\right)  $. We
perform infinite-depth operations \cite{arx2007wildeEAQCC}\ corresponding to
the entries in $\Gamma\left(  D\right)  $. The stabilizer matrix becomes%
\[
\left[  \left.
\begin{array}
[c]{ccc}%
I & \Gamma^{-1}\left(  D^{-1}\right)   & 0\\
0 & A\left(  D\right)   & Z_{2,N}^{\prime}\left(  D\right)
\end{array}
\right\vert
\begin{array}
[c]{ccc}%
0 & 0 & 0\\
I & \Gamma\left(  D\right)   & X_{2,N}^{\prime}\left(  D\right)
\end{array}
\right]  ,
\]
and the information-qubit matrix becomes%
\[
\left[  \left.
\begin{array}
[c]{ccc}%
0 & Z_{2,N}^{\prime\prime}\left(  D\right)  \Gamma^{-1}\left(  D^{-1}\right)
& 0\\
0 & X_{2,N}^{\prime\prime}\left(  D\right)  \Gamma^{-1}\left(  D^{-1}\right)
& I
\end{array}
\right\vert
\begin{array}
[c]{ccc}%
0 & 0 & I\\
0 & 0 & 0
\end{array}
\right]  .
\]
The above stabilizer matrix is equivalent to the desired one
in\ (\ref{eq:desired-check-matrix})\ by several row operations. We premultiply
the first set of rows by $\Gamma\left(  D^{-1}\right)  $ and multiply the
second set of rows by $\Gamma^{-1}\left(  D\right)  $. We can also use the
resulting identity matrix in the first set of rows to perform row operations
from the first set of rows to the second set of rows to realize the matrix
$X_{1}^{\prime}\left(  D\right)  $. The operators that Bob would really
measure need to have finite weight so he would measure the operators
corresponding to the entries in the following stabilizer matrix:%
\begin{equation}
\left[  \left.
\begin{array}
[c]{ccc}%
\Gamma\left(  D^{-1}\right)   & I & 0\\
0 & A\left(  D\right)   & Z_{2,N}^{\prime}\left(  D\right)
\end{array}
\right\vert
\begin{array}
[c]{ccc}%
0 & 0 & 0\\
I & \Gamma\left(  D\right)   & X_{2,N}^{\prime}\left(  D\right)
\end{array}
\right]  .\label{eq:end-encode}%
\end{equation}
We are done with the encoding algorithm. Alice begins with a set of ebits and
performs the encoding operations detailed in (\ref{eq:start-encode}%
-\ref{eq:end-encode}) and then performs the finite-depth operations detailed
in (\ref{eq:start-finite-depth-encode}-\ref{eq:desired-check-matrix})\ in
reverse order.
We now detail the steps of the decoding algorithm. Perform CNOT gates
corresponding to the entries in $X_{2,N}^{\prime}\left(  D\right)  $ from the
first set of $c$ qubits to the last set of $k+c$ qubits. The stabilizer matrix
becomes%
\[
\left[  \left.
\begin{array}
[c]{ccc}%
I & \Gamma^{-1}\left(  D^{-1}\right)  & 0\\
B\left(  D\right)  & A\left(  D\right)  & Z_{2,N}^{\prime}\left(  D\right)
\end{array}
\right\vert
\begin{array}
[c]{ccc}%
0 & 0 & 0\\
I & \Gamma\left(  D\right)  & 0
\end{array}
\right]  ,
\]
where $B\left(  D\right)  \equiv Z_{2,N}^{\prime}\left(  D\right)
X_{2,N}^{\prime\prime}\left(  D\right)  $. The information-qubit matrix
becomes%
\[
\left[  \left.
\begin{array}
[c]{ccc}%
0 & Z_{2,N}^{\prime\prime}\left(  D\right)  \Gamma^{-1}\left(  D^{-1}\right)
& 0\\
X_{2,N}^{\prime\prime}\left(  D\right)  & X_{2,N}^{\prime\prime}\left(
D\right)  \Gamma^{-1}\left(  D^{-1}\right)  & I
\end{array}
\right\vert
\begin{array}
[c]{ccc}%
0 & 0 & I\\
0 & 0 & 0
\end{array}
\right]  .
\]
Perform Hadamard gates on the last set of $k+c$ qubits. The stabilizer matrix
becomes%
\[
\left[  \left.
\begin{array}
[c]{ccc}%
I & \Gamma^{-1}\left(  D^{-1}\right)  & 0\\
B\left(  D\right)  & A\left(  D\right)  & 0
\end{array}
\right\vert
\begin{array}
[c]{ccc}%
0 & 0 & 0\\
I & \Gamma\left(  D\right)  & Z_{2,N}^{\prime}\left(  D\right)
\end{array}
\right]  ,
\]
and the information-qubit matrix becomes%
\[
\left[  \left.
\begin{array}
[c]{ccc}%
0 & Z_{2,N}^{\prime\prime}\left(  D\right)  \Gamma^{-1}\left(  D^{-1}\right)
& I\\
X_{2,N}^{\prime\prime}\left(  D\right)  & X_{2,N}^{\prime\prime}\left(
D\right)  \Gamma^{-1}\left(  D^{-1}\right)  & 0
\end{array}
\right\vert
\begin{array}
[c]{ccc}%
0 & 0 & 0\\
0 & 0 & I
\end{array}
\right]  .
\]
Perform CNOT gates from the first set of $c$ qubits to the last set of $k+c$
qubits. These CNOT\ gates correspond to the entries in $Z_{2,N}^{\prime
}\left(  D\right)  $. The stabilizer matrix becomes%
\[
\left[  \left.
\begin{array}
[c]{ccc}%
I & \Gamma^{-1}\left(  D^{-1}\right)  & 0\\
B\left(  D\right)  & A\left(  D\right)  & 0
\end{array}
\right\vert
\begin{array}
[c]{ccc}%
0 & 0 & 0\\
I & \Gamma\left(  D\right)  & 0
\end{array}
\right]  ,
\]
and the information-qubit matrix becomes%
\[
\left[  \left.
\begin{array}
[c]{ccc}%
Z_{2,N}^{\prime\prime}\left(  D\right)  & Z_{2,N}^{\prime\prime}\left(
D\right)  \Gamma^{-1}\left(  D^{-1}\right)  & I\\
X_{2,N}^{\prime\prime}\left(  D\right)  & X_{2,N}^{\prime\prime}\left(
D\right)  \Gamma^{-1}\left(  D^{-1}\right)  & 0
\end{array}
\right\vert
\begin{array}
[c]{ccc}%
0 & 0 & 0\\
0 & 0 & I
\end{array}
\right]  .
\]
Row operations from the first set of rows of the stabilizer to each of the two
sets of rows in the information-qubit matrix reduce the information-qubit
matrix to the following form:%
\[
\left[  \left.
\begin{array}
[c]{ccc}%
0 & 0 & I\\
0 & 0 & 0
\end{array}
\right\vert
\begin{array}
[c]{ccc}%
0 & 0 & 0\\
0 & 0 & I
\end{array}
\right]  .
\]
Then we perform the finite-depth operations detailed in
(\ref{eq:start-finite-depth-encode}-\ref{eq:desired-check-matrix}). We have
now finished the algorithm for the decoding circuit because the logical
operators for the information qubits appear in their original form.
\end{proof}

\subsection{Discussion}

Similar practical issues arise in these circuits as we discussed previously in
Ref.~\cite{arx2007wildeEAQCC}. Encoding circuits with infinite-depth
operations operations are acceptable if we assume that noiseless encoding is
possible. Otherwise, infinite-depth operations could lead to catastrophic
propagation of uncorrected errors. Noiseless encoding is difficult to achieve
in practice but we may be able to come close to it by concatenation of codes
at the encoder.

There is a dichotomy of these codes similar to that in Ref.
\cite{arx2007wildeEAQCC}. Some of the codes may have a simpler form in which
the encoding circuit requires finite-depth operations only.
Ref.~\cite{arx2008wildeUQCC}\ gives an example of a code with this structure.
These codes fall within the first class of codes discussed in
Ref.~\cite{arx2007wildeEAQCC} and will be more useful in practice because they
propagate errors in the encoding circuit to a finite number of qubits only.
The remaining codes that do not have this structure fall within the second
class of codes whose encoding circuits have both finite-depth and
infinite-depth operations and whose decoding circuits have finite-depth
operations only.

\subsection{Importing Classical Convolutional Codes over $GF\left(  4\right)
$}

One benefit of the new entanglement-assisted quantum convolutional codes is
that we can produce one from an arbitrary classical convolutional code over
$GF\left(  4\right)  $. The error-correcting properties of the classical
convolutional code translate to the resulting quantum convolutional code. It
is less clear how the rate translates because we use the expansion technique.
We know that the term $\left(  2k-n\right)  /n$ lower bounds the
\textquotedblleft entanglement-assisted\textquotedblright\ rate
\cite{arx2007wildeEAQCC}\ where $n$ and $k$ are the parameters from the
imported classical code. The rate should get a significant boost from
entanglement---the rate boosts by the number of ebits that the code requires.

The construction for importing an $\left[  n,k\right]  $ classical
convolutional code over $GF\left(  4\right)  $ is as follows. Suppose the
check matrix for the classical code is an $n-k\times n$-dimensional matrix
$H\left(  D\right)  $ whose entries are polynomials over $GF\left(  4\right)
$. We construct the quantum check matrix $\tilde{H}\left(  D\right)  $
according to the following formula:%
\[
\tilde{H}\left(  D\right)  =\gamma\left(  \left[
\begin{array}
[c]{c}%
\omega H\left(  D\right) \\
\bar{\omega}H\left(  D\right)
\end{array}
\right]  \right)  ,
\]
where $\gamma$ denotes the isomorphism between elements of $GF(4)$ and
symplectic binary vectors that represent Pauli matrices. Specifically,
$\gamma^{-1}(h)=\omega h_{x}+\bar{\omega}h_{z}$ where $h$ is a symplectic
binary vector and $h_{x}$ and $h_{z}$ denote its \textquotedblleft
X\textquotedblright\ and \textquotedblleft Z\textquotedblright\ parts
respectively. We use this construction for the example in the next section.

\section{Example}

\label{sec:examples}We take the convolutional generators from
Ref.~\cite{arx2007wildeCED} as our example. Ref.~\cite{science2006brun}
originally used these generators in a block code. Consider the following
generator of a check matrix for a classical convolutional code over GF$\left(
4\right)  $:%
\begin{equation}
\left(  \cdots|0000|1\bar{\omega}10|1101|0000|\cdots\right)  .
\end{equation}
The generators of the generator matrix for this code are as follows:
\begin{align}
&  \left(  \cdots|0000|1011|0000|0000|\cdots\right)  ,\\
&  \left(  \cdots|0000|1001|1010|0000|\cdots\right)  ,\\
&  \left(  \cdots|0000|01\bar{\omega}1|0000|0000|\cdots\right)  ,
\end{align}
because they are all in the null space of the above check matrix generator.
The distance of this classical convolutional code is three because the first
generator of the above three has the minimal weight of three. The
entanglement-assisted quantum convolutional code we produce below inherits
this distance.

We produce two quantum convolutional generators by mutliplying the above
generator by $\omega$ and $\bar{\omega}$ and applying the following map:%
\begin{equation}%
\begin{tabular}
[c]{l|l}\hline
$GF\left(  4\right)  $ & $\Pi$\\\hline
$0$ & $I$\\
$\omega$ & $X$\\
$1$ & $Y$\\
$\bar{\omega}$ & $Z$\\\hline
\end{tabular}
\ \ \ \ \ ,
\end{equation}
where $\Pi$ denotes the group of Pauli matrices for one qubit. The resulting
quantum convolutional generators are as follows:%
\begin{align}
&  \left(  \cdots|IIII|ZXZI|ZZIZ|IIII|\cdots\right)  ,\nonumber\\
&  \left(  \cdots|IIII|XYXI|XXIX|IIII|\cdots\right)  .
\end{align}
It is not a CSS\ code because the original classical generator has support on
the element $\bar{\omega}$. These generators have the following representation
in the polynomial formalism:%
\[
\left[  \left.
\begin{array}
[c]{cccc}%
1+D & D & 1 & D\\
0 & 1 & 0 & 0
\end{array}
\right\vert
\begin{array}
[c]{cccc}%
0 & 1 & 0 & 0\\
1+D & 1+D & 1 & D
\end{array}
\right]  .
\]
The shifted symplectic product matrix $\Omega\left(  D\right)  $ for the above
code is as follows:%
\[
\Omega\left(  D\right)  =%
\begin{bmatrix}
D+D^{-1} & D^{-1}\\
D & D+D^{-1}%
\end{bmatrix}
.
\]
The above matrix is not reducible to the standard form by any row operations.
We therefore expand the code by a factor of two to give four generators with a
frame size of eight. The \textquotedblleft Z\textquotedblright\ matrix
$Z\left(  D\right)  $ of the two-expanded check matrix is as follows:%
\[
Z\left(  D\right)  =%
\begin{bmatrix}
1 & 0 & 1 & 0 & 1 & 1 & 0 & 1\\
0 & 1 & 0 & 0 & 0 & 0 & 0 & 0\\
D & D & 0 & D & 1 & 0 & 1 & 0\\
0 & 0 & 0 & 0 & 0 & 1 & 0 & 0
\end{bmatrix}
,
\]
and the \textquotedblleft X\textquotedblright\ matrix $X\left(  D\right)
$\ of the two-expanded check matrix is as follows:%
\[
X\left(  D\right)  =%
\begin{bmatrix}
0 & 1 & 0 & 0 & 0 & 0 & 0 & 0\\
1 & 1 & 1 & 0 & 1 & 1 & 0 & 1\\
0 & 0 & 0 & 0 & 0 & 1 & 0 & 0\\
D & D & 0 & D & 1 & 1 & 1 & 0
\end{bmatrix}
.
\]
The shifted symplectic product matrix $\Omega_{2}\left(  D\right)  $ of the
two-expanded check matrix is as follows:%
\[
\Omega_{2}\left(  D\right)  =%
\begin{bmatrix}
0 & 0 & 1+D^{-1} & D^{-1}\\
0 & 0 & 1 & 1+D^{-1}\\
1+D & 1 & 0 & 0\\
D & 1+D & 0 & 0
\end{bmatrix}
.
\]
We proceed with the Gram-Schmidt procedure because this matrix satisfies its
initial requirements. We swap generators two and three to be the first and
second generators of the check matrix because they have the commutation
relations of half of an ebit. The shifted symplectic product matrix becomes%
\[%
\begin{bmatrix}
0 & 1 & 0 & 1+D^{-1}\\
1 & 0 & 1+D & 0\\
0 & 1+D^{-1} & 0 & D^{-1}\\
1+D & 0 & D & 0
\end{bmatrix}
.
\]
Multiply generator two by $1+D$ and add to generator four. Multiply generator
one by $1+D^{-1}$ and add to generator three. The shifted symplectic matrix
becomes%
\[%
\begin{bmatrix}
0 & 1 & 0 & 0\\
1 & 0 & 0 & 0\\
0 & 0 & 0 & 1+D^{-1}+D^{-2}\\
0 & 0 & 1+D+D^{2} & 0
\end{bmatrix}
.
\]
We finally divide generator four by $1+D+D^{2}$ and the shifted symplectic
product matrix then becomes%
\[%
%TCIMACRO{\dbigoplus \limits_{i=1}^{2}}%
%BeginExpansion
{\displaystyle\bigoplus\limits_{i=1}^{2}}
%EndExpansion
J\text{,}%
\]
so that it has the commutation relations of halves of two ebits. The check
matrix resulting from these operations is as follows:%
\begin{equation}
H_{2}\left(  D\right)  =\left[  \left.
\begin{array}
[c]{c}%
Z_{2}\left(  D\right)
\end{array}
\right\vert
\begin{array}
[c]{c}%
X_{2}\left(  D\right)
\end{array}
\right]  , \label{eq:example-code}%
\end{equation}
where%
\[
Z_{2}^{T}\left(  D\right)  =\left[
\begin{array}
[c]{cccc}%
0 & D & 1 & \frac{D^{2}+D}{D^{2}+D+1}\\
1 & D & \frac{1}{D}+1 & \frac{D^{2}+D}{D^{2}+D+1}\\
0 & 0 & 1 & 0\\
0 & D & 0 & \frac{D^{2}+D}{D^{2}+D+1}\\
0 & 1 & 1 & \frac{D+1}{D^{2}+D+1}\\
0 & 0 & 1 & \frac{1}{D^{2}+D+1}\\
0 & 1 & 0 & \frac{D+1}{D^{2}+D+1}\\
0 & 0 & 1 & 0
\end{array}
\right]  ,\ \ \ \ \ \ \ \ X_{2}^{T}\left(  D\right)  =\left[
\begin{array}
[c]{cccc}%
1 & 0 & 1+D^{-1} & \frac{D}{D^{2}+D+1}\\
1 & 0 & \frac{1}{D} & \frac{D}{D^{2}+D+1}\\
1 & 0 & 1+D^{-1} & 0\\
0 & 0 & 0 & \frac{D}{D^{2}+D+1}\\
1 & 0 & 1+D^{-1} & \frac{1}{D^{2}+D+1}\\
1 & 1 & 1+D^{-1} & \frac{D}{D^{2}+D+1}\\
0 & 0 & 0 & \frac{1}{D^{2}+D+1}\\
1 & 0 & 1+D^{-1} & 0
\end{array}
\right]  .
\]
(We give the transposition of the above two matrices because they otherwise
would not fit in the space above.) The error-correcting properties of the
above check matrix are equivalent to the error-correcting properties of the
original two generators.

This code sends six information qubits and consumes two ebits per eight
channel uses. The rate pair for this code is therefore $\left(
3/4,1/4\right)  $.

We can now apply the algorithm in Theorem~\ref{thm:main} to determine the
encoding and decoding circuits for this code. The encoding circuit begins from
a set of two ebits and eight information qubits per frame with the following
stabilizer matrix:%
\[
H_{0}\left(  D\right)  =\left[  \left.
\begin{array}
[c]{c}%
Z_{0}\left(  D\right)
\end{array}
\right\vert
\begin{array}
[c]{c}%
X_{0}\left(  D\right)
\end{array}
\right]  ,
\]
where%
\[
Z_{0}\left(  D\right)  =\left[
\begin{array}
[c]{cccccccccc}%
1 & 0 & 1 & 0 & 0 & 0 & 0 & 0 & 0 & 0\\
0 & 1 & 0 & 1 & 0 & 0 & 0 & 0 & 0 & 0\\
0 & 0 & 0 & 0 & 0 & 0 & 0 & 0 & 0 & 0\\
0 & 0 & 0 & 0 & 0 & 0 & 0 & 0 & 0 & 0
\end{array}
\right]  ,\ \ \ \ \ \ \ \ X_{0}\left(  D\right)  =\left[
\begin{array}
[c]{cccccccccc}%
0 & 0 & 0 & 0 & 0 & 0 & 0 & 0 & 0 & 0\\
0 & 0 & 0 & 0 & 0 & 0 & 0 & 0 & 0 & 0\\
1 & 0 & 1 & 0 & 0 & 0 & 0 & 0 & 0 & 0\\
0 & 1 & 0 & 1 & 0 & 0 & 0 & 0 & 0 & 0
\end{array}
\right]  .
\]
We label the eight qubits on the right side of each matrix above as
$1,\ldots,8$. We label Bob's two qubits on the left as $B1$ and $B2$. Perform
the following finite-depth operations (in order from left to right and then
top to bottom):%
\begin{align*}
&  C\left(  1,4,D+D^{2}\right)  C\left(  1,5,1+D^{2}\right)  C\left(
1,6,1\right)  C\left(  1,7,1+D\right) \\
&  C\left(  2,4,D\right)  C\left(  2,5,1+D\right)  C\left(  2,6,1\right) \\
&  H\left(  3,\ldots,8\right)  C\left(  1,4,D+D^{2}+D^{4}\right)  C\left(
1,5,D^{2}\right) \\
&  C\left(  1,6,1+D\right)  C\left(  1,7,D^{2}\right)  C\left(  2,4,D+D^{2}%
\right) \\
&  C\left(  2,5,1+D\right)  C\left(  2,6,1\right)  C\left(  2,7,1+D\right)  ,
\end{align*}
where we use the notation $C\left(  q_{1},q_{2},f\left(  D\right)  \right)  $
to represent a finite-depth CNOT\ gate from qubit one to qubit two that
implements the polynomial $f\left(  D\right)  $, $H\left(  q_{i},\ldots
,q_{j}\right)  $ is a sequence of Hadamard gates applied to qubits $q_{i}$
through $q_{j}$ in each frame, $P\left(  q\right)  $ is a phase gate applied
to qubit $q$ in each frame, and $C\left(  q,1/f\left(  D\right)  \right)  $ is
an infinite-depth CNOT\ gate implementing the rational polynomial $1/f\left(
D\right)  $. Alice performs the following infinite-depth operations:%
\begin{align*}
&  H\left(  1,2\right)  C\left(  1,\frac{1}{1+D^{-1}+D^{-2}}\right) \\
&  C\left(  2,\frac{1}{1+D^{-1}+D^{-2}}\right)  H\left(  1,2\right)  .
\end{align*}
She then finishes the encoding circuit with the following finite-depth
operations:%
\begin{align}
&  H\left(  1,2\right)  C\left(  2,3,1\right)  C\left(  2,5,1\right)  C\left(
2,6,1\right)  C\left(  2,8,1\right) \label{eq:example-finite-depth}\\
&  P\left(  2\right)  H\left(  3,\ldots,8\right)  S\left(  2,3\right)
\nonumber\\
&  C\left(  1,2,1\right)  C\left(  1,3,1\right)  C\left(  1,5,1\right)
C\left(  1,6,1\right)  C\left(  1,8,1\right)  P\left(  2\right)  .\nonumber
\end{align}
The code she encodes has equivalent error-correcting properties to the check
matrix in (\ref{eq:example-code}).

Bob performs the following operations in the decoding circuit. He first
performs the operations in (\ref{eq:example-finite-depth})\ in reverse order.
He then performs the following finite-depth operations:%
\begin{align*}
&  C\left(  B1,4,D+D^{2}+D^{4}\right)  C\left(  B1,5,D^{2}\right) \\
&  C\left(  B1,6,1+D\right)  C\left(  B1,7,D^{2}\right)  C\left(
B2,4,D+D^{2}\right) \\
&  C\left(  B2,5,1+D\right)  C\left(  B2,6,1\right)  C\left(  B2,7,1+D\right)
\\
&  H\left(  3,\ldots,8\right) \\
&  C\left(  B1,4,D+D^{2}\right)  C\left(  B1,5,1+D^{2}\right)  C\left(
B1,6,1\right) \\
&  C\left(  B1,7,1+D\right)  C\left(  B2,4,D\right) \\
&  C\left(  B2,5,1+D\right)  C\left(  B2,6,1\right)  .
\end{align*}
The information qubits then appear at the output of this online decoding circuit.

\section{Conclusion and Current Work}

\label{sec:conclusion}There are several differences between the methods used
for general, non-CSS\ codes discussed in this article and the CSS\ codes used
in our previous article~\cite{arx2007wildeEAQCC}. It was more straightforward
to determine how to use ebits efficiently in CSS entanglement-assisted quantum
convolutional codes, but we have had to introduce the expansion technique in
Section~\ref{sec:expand}\ in order to determine how to use ebits efficiently
for codes in this article. There was also no need for an explicit Gram-Schmidt
orthogonalization procedure in Ref.~\cite{arx2007wildeEAQCC}. The Smith
algorithm implicitly produced the row operations necessary for symplectic orthogonalization.

We do have some methods that do not require expansion of a check matrix or an
explicit Gram-Schmidt procedure~\cite{PhysRevA.79.032313}, but these methods
do not make efficient use of entanglement and have a lower rate of noiseless
qubit channel simulation and higher rate of entanglement consumption than the
codes discussed in this article. Nonetheless, we have determined ways to make
these other codes more useful by encoding classical information in the extra
entanglement with a superdense-coding-like effect \cite{PhysRevLett.69.2881}.
These other codes are grandfather codes in the sense of
Refs.~\cite{arx2008wildeUQCC,arx2008kremsky} because they consume entanglement
to send both quantum and classical information.

One negative implication of the expansion of a code is that the expanded code
requires more qubits per frame. The expanded code then requires a larger
buffer at both the sender's and receiver local stations. The increased buffer
will be a concern right now because it is difficult to build large quantum
memories. This issue will become less of a concern as quantum technology
advances. The entanglement-inefficient codes mentioned in the previous
paragraph have the advantage that they do not require expansion and thus
require smaller buffers. It therefore should be of interest to find solutions
in between the entanglement-efficient codes discussed in this article and the
entanglement-inefficient codes discussed in the previous paragraph.

Some outstanding issues remain. We have not proven the convergence of the
polynomial symplectic Gram-Schmidt orthogonalization procedure and have
instead provided a practical stopping condition. We have a conjecture for how
to proceed with proving convergence. Suppose that we would like to construct a
code consisting of one generator that does not commute with shifts of itself.
We have found for many examples that the correct expansion factor for the
generator is equal to the period of the inverse polynomial of the generator's
shifted symplectic product. We do not have a proof that this factor is the
correct one and we are not sure what the expansion factor should be when we
would like to construct a code starting from more than one generator. The
conjecture about optimal entanglement use in general entanglement-assisted
quantum convolutional codes from Ref.~\cite{arx2008wildeOEA}\ also remains an
open question. The conjecture is that the optimal number of ebits required per
frame is equal to the following expression:%
\[
\text{rank}\left(  Z\left(  D\right)  X^{T}\left(  D^{-1}\right)  +X\left(
D\right)  Z^{T}\left(  D^{-1}\right)  \right)  /2,
\]
where $Z\left(  D\right)  $ and $X\left(  D\right)  $ are the respective
\textquotedblleft Z\textquotedblright\ and \textquotedblleft
X\textquotedblright\ matrices for a given set of quantum convolutional
generators. It is clear that this formula holds after we have expanded the
original set of generators. The proof technique follows from the proof
technique outlined in Ref.~\cite{arx2008wildeOEA}. But we are not sure how to
apply this formula to an initial set of unexpanded generators.

The techniques developed in this article represent a useful way for encoding
quantum information. The next step should be to combine the theory in this
article with Poulin \textit{et al}.'s recent theory of quantum serial-turbo
coding~\cite{arx2007poulin}.

MMW\ acknowledges support from NSF Grant No.~CCF-0545845,\ and
TAB\ acknowledges support from NSF Grant No.~CCF-0448658 and No.~CCF-0830801.


\begin{thebibliography}{99}                                                                                               %


\bibitem {PhysRevLett.91.177902}Harold Ollivier and Jean-Pierre Tillich.
\newblock Description of a quantum convolutional code.
\newblock {\em Physical Review Letters}, 91(17):177902, October 2003.

\bibitem {arxiv2004olliv}Harold Ollivier and Jean-Pierre Tillich.
\newblock Quantum convolutional codes: Fundamentals.
\newblock {\em arXiv:quant-ph/0401134}, 2004.

\bibitem {isit2006grassl}Markus Grassl and Martin R\"{o}tteler.
\newblock Noncatastrophic encoders and encoder inverses for quantum
convolutional codes. \newblock In \emph{IEEE International Symposium on
Information Theory (quant-ph/0602129)}, 2006.

\bibitem {ieee2006grassl}Markus Grassl and Martin R\"{o}tteler.
\newblock Quantum convolutional codes: Encoders and structural properties.
\newblock In \emph{Forty-Fourth Annual Allerton Conference}, 2006.

\bibitem {ieee2007grassl}Markus Grassl and Martin R\"{o}tteler.
\newblock Constructions of quantum convolutional codes. \newblock In
\emph{IEEE International Symposium on Information Theory}, 2007.

\bibitem {isit2005forney}G.~David Forney and Saikat Guha. \newblock Simple
rate-1/3 convolutional and tail-biting quantum error-correcting codes.
\newblock In \emph{IEEE International Symposium on Information Theory
(arXiv:quant-ph/0501099)}, 2005.

\bibitem {ieee2007forney}G.~David Forney, Markus Grassl, and Saikat Guha.
\newblock Convolutional and tail-biting quantum error-correcting codes.
\newblock {\em IEEE Transactions on Information Theory}, 53:865--880, 2007.

\bibitem {cwit2007aly}Salah~A. Aly, Markus Grassl, Andreas Klappenecker,
Martin Roetteler, and Pradeep~Kiran Sarvepalli. \newblock Quantum
convolutional BCH codes. \newblock In \emph{10th Canadian Workshop on
Information Theory (arXiv:quant-ph/0703113)}, pages 180--183, 2007.

\bibitem {arx2007aly}Salah~A. Aly, Andreas Klappenecker, and Pradeep~Kiran
Sarvepalli. \newblock Quantum convolutional codes derived from Reed-Solomon
and Reed-Muller codes. \newblock {\em arXiv:quant-ph/0701037}, 2007.

\bibitem {arx2007wildeCED}Mark~M. Wilde, Hari Krovi, and Todd~A. Brun.
\newblock Convolutional entanglement distillation. \newblock To appear in the
\emph{International Symposium on Information Theory}, arXiv:0708.3699, Austin,
Texas, USA June 2010.

\bibitem {arx2007wildeEAQCC}Mark~M. Wilde and Todd~A. Brun.
\newblock Entanglement-assisted quantum convolutional coding. \newblock To
appear in \emph{Physical Review A}, arXiv:0712.2223, 2010.

\bibitem {arx2008wildeUQCC}Mark~M. Wilde and Todd~A. Brun. \newblock Unified
quantum convolutional coding. \newblock In \emph{IEEE International Symposium
on Information Theory (arXiv:0801.0821)}, July 2008.

\bibitem {PhysRevA.55.1613}Seth Lloyd. \newblock Capacity of the noisy quantum
channel. \newblock {\em Physical Review A}, 55(3):1613--1622, March 1997.

\bibitem {capacity2002shor}Peter~W. Shor. \newblock The quantum channel
capacity and coherent information. \newblock In \emph{Lecture Notes, MSRI
Workshop on Quantum Computation}, 2002.

\bibitem {ieee2005dev}Igor Devetak. \newblock The private classical capacity
and quantum capacity of a quantum channel.
\newblock {\em IEEE Transactions on Information Theory}, 51:44--55, January 2005.

\bibitem {qcap2008first}Patrick Hayden, Michal Horodecki, Andreas Winter, and
Jon Yard. \newblock A decoupling approach to the quantum capacity.
\newblock {\em Open Systems \& Information Dynamics}, 15:7 -- 19, March 2008.

\bibitem {qcap2008second}Rochus Klesse. \newblock A random coding based proof
for the quantum coding theorem.
\newblock {\em Open Systems \& Information Dynamics}, 15:21--45, March 2008.

\bibitem {qcap2008third}Michal Horodecki, Seth Lloyd, and Andreas Winter.
\newblock Quantum coding theorem from privacy and distinguishability.
\newblock {\em Open Systems \& Information Dynamics}, 15:47--69, March 2008.

\bibitem {qcap2008fourth}Patrick Hayden, Peter~W. Shor, and Andreas Winter.
\newblock Random quantum codes from {Gaussian} ensembles and an uncertainty
relation. \newblock {\em Open Systems \& Information Dynamics}, 15:71--89,
March 2008.

\bibitem {arx2007poulin}David Poulin, Jean-Pierre Tillich, and Harold
Ollivier. \newblock Quantum serial turbo-codes.
\newblock {\em IEEE Transactions on Information Theory}, 55(6):2776--2798,
June 2009.

\bibitem {ieee1998calderbank}A.R. Calderbank, E.M. Rains, P.W. Shor, and
N.J.A. Sloane. \newblock Quantum error correction via codes over {GF(4)}.
\newblock {\em IEEE Transactions on Information Theory}, 44:1369--1387, 1998.

\bibitem {book2000mikeandike}Michael~A. Nielsen and Isaac~L. Chuang.
\newblock {\em Quantum Computation and Quantum Information}.
\newblock Cambridge University Press, 2000.

\bibitem {science2006brun}Todd~A. Brun, Igor Devetak, and Min-Hsiu Hsieh.
\newblock Correcting quantum errors with entanglement.
\newblock {\em Science}, 314(5798):436--439, October 2006.

\bibitem {arx2006brun}Todd~A. Brun, Igor Devetak, and Min-Hsiu Hsieh.
\newblock Catalytic quantum error correction.
\newblock {\em arXiv:quant-ph/0608027}, August 2006.

\bibitem {grassl2006itw}Markus Grassl. \newblock Convolutional and block
quantum error-correcting codes. \newblock In \emph{IEEE Information Theory
Workshop}, pages 144--148, Chengdu, October 2006.

\bibitem {book1999conv}Rolf Johannesson and Kamil~Sh. Zigangirov.
\newblock {\em Fundamentals of Convolutional Coding}. \newblock Wiley-IEEE
Press, 1999.

\bibitem {thesis97gottesman}Daniel Gottesman.
\newblock {\em Stabilizer Codes and Quantum Error Correction}. \newblock PhD
thesis, California Institue of Technology, 1997.

\bibitem {prep2007shaw}Bilal Shaw, Mark~M. Wilde, Ognyan Oreshkov, Isaac
Kremsky, and Daniel Lidar. \newblock Encoding one logical qubit into six
physical qubits. \newblock {\em Physical Review A}, 78:012337, 2008.

\bibitem {arx2008wildeOEA}Mark~M. Wilde and Todd~A. Brun. \newblock Optimal
entanglement formulas for entanglement-assisted quantum coding.
\newblock {\em Physical Review A}, 77:064302, 2008.

\bibitem {PhysRevLett.69.2881}Charles~H. Bennett and Stephen~J. Wiesner.
\newblock Communication via one- and two-particle operators on
Einstein-Podolsky-Rosen states. \newblock {\em Physical Review Letters},
69(20):2881--2884, November 1992.

\bibitem {PhysRevA.79.032313}Mark~M.~Wilde and Todd~A. Brun. \newblock Extra
shared entanglement reduces memory demand in quantum convolutional coding.
\newblock {\em Physical Review A}, 79(3):032313, March 2009.

\bibitem {arx2008kremsky}Isaac Kremsky, Min-Hsiu Hsieh, and Todd~A. Brun.
\newblock Classical enhancement of quantum-error-correcting codes.
\newblock {\em Physical Review A}, 78(1):012341, July 2008.
\end{thebibliography}
\end{document}